\documentclass[twocolumn,tighten,times,astrosymb,floatfix]{aastex631}

\usepackage{amsmath}

\begin{document}

\title{A \textit{JWST} MIRI MRS View of the $\eta$ Tel Debris Disk and its Brown Dwarf Companion}

\correspondingauthor{Yiwei Chai}
\email{mchai3@jhu.edu}

\author[0009-0008-5865-5831]{Yiwei Chai}
\affiliation{William H. Miller III Department of Physics and Astronomy, Johns Hopkins University, 3400 N. Charles Street, Baltimore, MD 21218, USA}

\author[0000-0002-8382-0447]{Christine H. Chen}
\affiliation{Space Telescope Science Institute, 3700 San Martin Drive, Baltimore, MD 21218, USA}
\affiliation{William H. Miller III Department of Physics and Astronomy, Johns Hopkins University, 3400 N. Charles Street, Baltimore, MD 21218, USA}

\author[0000-0002-5885-5779]{Kadin Worthen}
\affiliation{William H. Miller III Department of Physics and Astronomy, Johns Hopkins University, 3400 N. Charles Street, Baltimore, MD 21218, USA}

\author[0009-0001-7058-8538]{Alexis Li}
\affiliation{William H. Miller III Department of Physics and Astronomy, Johns Hopkins University, 3400 N. Charles Street, Baltimore, MD 21218, USA}

\author[0000-0003-4623-1165]{Antranik A. Sefilian}
\affiliation{Astrophysikalisches Institut und Universit{\"a}tssternwarte, Friedrich-Schiller-Universit\"at Jena, Schillerg\"a{\ss}chen~2--3, D-07745 Jena, Germany}
\affiliation{Center for Advanced Mathematical Sciences, American University of Beirut, P.O. Box 11-0236, Riad El-Solh, Beirut 11097 2020, Lebanon}

\author[0000-0001-6396-8439]{William Balmer}
\affiliation{William H. Miller III Department of Physics and Astronomy, Johns Hopkins University, 3400 N. Charles Street, Baltimore, MD 21218, USA}

\author[0000-0003-4653-6161]{Dean C. Hines}
\affiliation{Space Telescope Science Institute, 3700 San Martin Drive, Baltimore, MD 21218, USA}

\author[0000-0002-9402-186X]{David R. Law}
\affiliation{Space Telescope Science Institute, 3700 San Martin Drive, Baltimore, MD 21218, USA}

\author[0000-0001-9855-8261]{B. A. Sargent}
\affiliation{Space Telescope Science Institute, 3700 San Martin Drive, Baltimore, MD 21218, USA}
\affiliation{William H. Miller III Department of Physics and Astronomy, Johns Hopkins University, 3400 N. Charles Street, Baltimore, MD 21218, USA}

\author{Mark Wyatt}
\affiliation{Institute of Astronomy, University of Cambridge, Madingley Road, Cambridge CB3 0HA, UK}

\author[0000-0001-9352-0248]{Cicero X. Lu}
\affiliation{Gemini Observatory/NSF’s NOIRLab, 670N. A’ohoku Place, Hilo, HI 96720, USA}

\author[0000-0002-3191-8151]{Marshall D. Perrin}
\affiliation{Space Telescope Science Institute, 3700 San Martin Drive, Baltimore, MD 21218, USA}

\author[0000-0002-4388-6417]{Isabel Rebollido}
\affiliation{Centro de Astrobiología (CAB), INTA-CSIC, Camino Bajo del Castillo s/n - Villafranca del Castillo, 28692 Villanueva de la Cañada, Madrid, Spain}

\author[0000-0003-4203-9715]{Emily Rickman}
\affiliation{European Space Agency (ESA), ESA Office, Space Telescope Science Institute, 3700 San Martin Drive, Baltimore, MD 21218, USA}

\author[0000-0003-4520-1044]{G.\ C.\ Sloan}
\affiliation{Space Telescope Science Institute, 3700 San Martin Drive,
  Baltimore, MD 21218, USA}
\affiliation{Department of Physics and Astronomy, University of North
  Carolina at Chapel Hill, Chapel Hill, NC 27599-3255, USA}

\begin{abstract}

We report \textit{JWST} MIRI MRS observations of the $\beta$ Pic moving group member, $\eta$ Tel A, along with its brown dwarf binary companion, $\eta$ Tel B. Following PSF subtraction, we recover the spatially resolved flux from the debris disk around $\eta$ Tel A, along with the position of the companion exterior to the disk. We present a new 5--26 $\mu m$ epoch of spectroscopy for the disk, in which we discover a 20 $\mu m$ silicate feature, and the first ever 11--21 $\mu m$ spectrum of $\eta$ Tel B, which indicates a bare photosphere. We derive a new epoch of relative astrometry for the companion, extending the baseline of measurements to 25 years, and find that it is currently located near the apocentre of an eccentric, long-period orbit. The companion's orbit is close enough to the disk that it should significantly perturb the planetesimals within it, resulting in a detectable mid-IR pericenter glow and near-alignment with the companion. Contrary to expectations, however, we find that the disk appears to be axisymmetric and potentially misaligned with the companion in the MIRI MRS data. We posit that this may be due to the presence of an additional, yet-undetected $\sim$0.7--30 $M_J$ planet orbiting interior to the disk with a semimajor axis of $\sim$3--19 au.

\end{abstract}




\section{Introduction}
\label{sec:intro}

Observations of planetary systems with multiple components offer a valuable opportunity to understand the interplay and mutual influence of objects within the system throughout its evolutionary timeline. Young systems with circumstellar disks (e.g. debris disks) in particular offer an exciting view into the sculpting of disk morphologies due to dynamical interactions in the system. Understanding these dynamically-induced disk structures can also aid in inferring the presence of yet-undetected planets within systems that have circumstellar disks \citep{Wyatt1999, Hughes2018, Wyatt2020}.

To date, however, there are only a few observed examples of young debris disk systems with wide-separation binary companions (e.g. HD 106906 \citep{Bailey2014, Rodet2017}). $\eta$ Telescopii (henceforth $\eta$ Tel), as a relatively young ($\sim$23 Myr; \cite{Mamajek2014}) debris disk--hosting triple system, therefore offers an interesting target for observational study.

The $\eta$ Tel system is located 49.5 pc away \citep{GAIA2023} within the $\beta$ Pictoris moving group, and consists of (1) $\eta$ Tel A, an A0V-type primary \citep{Houk1975}, (2) $\eta$ Tel B, a M7/8-type brown dwarf companion at a separation of 4" \citep{Lowrance2000}, and (3) HD 181327, a F6-type co-moving star at a separation of 7'. The primary is host to an edge-on, North-South aligned debris disk extending to at least 24 au in the mid-IR \citep{Smith2009}. Interestingly, HD 181327 is also host to a well-studied debris disk, albeit one that is face-on \citep{Schneider2006, Marino2016, Milli2023}.

The debris disk around $\eta$ Tel A was first identified based on \textit{IRAS} measurements indicating an excess in emission at 12, 25, and 60 $\mu m$ \citep{Backman1993}, for which a dust optical depth was calculated to be $\tau=L_{IR}/L_*\approx3.5\times10^{-4}$ \citep{Lowrance2000}. A 2004 \textit{Spitzer} IRS observation revealed a largely featureless spectrum from 5--35 $\mu m$, with the exception of a possible 10 $\mu m$ silicate feature suggesting the presence of large grains \citep{Chen2006}. Additionally, the excess emission was found to be best fit by two different temperatures of dust: a `warm' 370 K component, and a `cool' 116 K component. However, from the spectrum alone, it was not possible to resolve degeneracies regarding the spatial structure of the dust (e.g. two dust populations at different locations versus two populations with different grain sizes at the same location). 18.3 $\mu m$ ground-based imaging with T-ReCs on Gemini South spatially resolved the outer component of the disk \citep{Smith2009}. Modelling of the T-ReCs disk images was consistent with a two-component disk structure comprising of an unresolved inner 'warm' component inwards of $\sim$4 au (as also inferred by \cite{Chen2006}), and a resolved `cool' component in the shape of a narrow ring centred at 24 au. 

High resolution optical spectroscopy of the $\eta$ Tel disk using FEROS detected Ca II K absorption lines at $\sim$-23 km s$^{-1}$ that were attributed to circumstellar gas \citep{Rebollido2018}. Far- and near-UV spectroscopy with \textit{HST} STIS likewise detected absorption features at -23 km s$^{-1}$ for multiple atomic lines, as well as features at -18 km s$^{-1}$ \citep{Youngblood2021}. The -23 km s$^{-1}$ and -18 km s$^{-1}$ components were respectively attributed to circumstellar and to interstellar gas. In particular, the blueshifting of the -23 km s$^{-1}$ absorption features with respect to the star's reference frame was interpreted to indicate gas outflow in a radiatively driven disk wind. However, subsequent work \citep{Iglesias2023} tested the posited circumstellar origin of the gas by comparing the $\eta$ Tel absorption features to those of HD 181327, HD 180575, and $\rho$ Tel, three stars with a similar line of sight. The absorption features at $\sim$-23 km s$^{-1}$ were found in the Ca II K lines of the three other stars, strongly implying that the $\eta$ Tel absorption lines attributed to circumstellar gas are instead more likely due to an interstellar cloud traversing $\eta$ Tel's line of sight. 

The brown dwarf companion, $\eta$ Tel B (a.k.a. HR 7329 B), was first discovered with \textit{HST} NICMOS coronography at a separation of 4" from the primary \citep{Lowrance2000}. \textit{HST} STIS spectroscopy indicated a spectral type of M7-8 (Lowrence et al. 2000), which was confirmed with H-band spectroscopy from VLT ISAAC \citep{Guenther2001}. Although initial attempts to show common proper motion from measurements of the companion's separation and position angle (PA) were inconclusive \citep{Guenther2001}, additional imaging observations from \textit{HST} NICMOS and VLT NACO across a baseline of 11 years between 1998 and 2009 were used to confirm $\eta$ Tel B's status as a comoving companion \citep{Neuhauser2011}, possibly detecting a small linear change in separation (2.91 $\pm$ 2.41 mas yr$^{-1}$) and finding no change in PA. It was suggested that this indicates the companion is currently located near apocentre of an inclined and/or eccentric orbit. Magnitude estimates for $\eta$ Tel B were also used to derive a mass of 20-50 $M_J$ from evolutionary tracks \citep{Neuhauser2011}. No additional companions up to 9" separation from the primary were detected in the 1998 \textit{HST} NICMOS and 2004--2008 VLT NACO H-band images. Later coronagraphic imaging with SPHERE/IRDIS from 2015--2017 likewise did not detect any satellites around the companion itself, placing an upper limit on potential satellites from 3 $M_J$ at 10 au to 1.6 $M_J$ at 33 au \citep{nogueira2024astrometric}.

Additionally, several attempts have been made to characterise the orbit of $\eta$ Tel B from its astrometric measurements. An analytical approach assuming a face-on, circular orbit gave a companion semi-major axis of $a=220^{214}_{-84}$ au, and a period of $\sim$2000 years \citep{Neuhauser2011}; this was refined given the existence (and thus stability) of the edge-on debris disk around $\eta$ Tel A, which allows for a potential constraint to be placed on the eccentricity of the companion's orbit. Assuming that $\eta$ Tel B's apocentre distance is indeed $r_\mathrm{max}\sim200$ au, and that it has sculpted the outer edge of the debris disk around $\eta$ Tel A to be $r_\mathrm{disk}\sim$ 24 au, $e=0.47$. This gives a semi-major axis $a=136$ au and an orbital period of $P\sim1000$ years \citep{Neuhauser2011}. \cite{Blunt2017} used the same 11 year baseline of astrometric measurements from \cite{Neuhauser2011} to perform an orbital fit using the OFTI (Orbits for the Impatient) algorithm, obtaining median orbital parameters of $a=192^{+240}_{-67}$ au, $P=1490^{+3350}_{-710}$ yr, $e=0.77^{+0.19}_{-0.43}$, and $i=86^{+10}_{-19}$ deg, with uncertainties at a 68\% (1-$\sigma$) confidence interval. Most recently, the orbit-fitting package \texttt{orvara} \citep{Brandt2021} was used to derive the companion's orbital parameters from 2015--2017 SPHERE/IRDIS observations combined with previous astrometric measurements over a baseline of 19 years \citep{nogueira2024astrometric}. This fit reported an inclination of $i=82^{+3}_{-4}$ degrees, a semi-major axis of $a=218^{+180}_{-41}$ au, and an eccentricity of $e=0.34\pm0.26$. While the orbital inclination has been fairly consistent across the literature, derivations of the companion's semi-major axis and eccentricity remain relatively poorly constrained. To date, the large uncertainties on these two parameters illustrate the challenge of characterising the orbit of long-period companions, for which astrometric observations may only cover a small fraction of the total orbital period.

In this paper, we present a new observation of the $\eta$ Tel system with \textit{JWST} MIRI MRS. \hyperref[sec:obsndata]{Section 2} details the observation and processing of the data. \hyperref[sec:Adisk]{Section 3} presents a new epoch of mid-IR spectroscopy for $\eta$ Tel A and the discovery of a 20 $\mu m$ silicate feature, dust modelling for the MRS spectrum, and our analysis of the spatially resolved disk. In \hyperref[sec:Bdisk]{Section 4}, we present the first 11-21 $\mu m$ spectrum for the brown dwarf companion, $\eta$ Tel B, finding that the object does not possess a mid-IR excess. In \hyperref[sec:Borb]{Section 5}, we discuss (1) our new epoch of astrometry for $\eta$ Tel B from MIRI MRS, which extends the baseline of measurements to 25 years and (2) our new orbital derivation for the companion. In \hyperref[sec:discussion]{Section 6}, we consider how dynamical interactions with $\eta$ Tel B are expected to impact the radial extent and symmetry of $\eta$ Tel A disk; we suggest that a yet-undetected planetary mass may explain the disagreement between our observations and the expected effects from the companion. We summarise our results and state our conclusions in \hyperref[sec:concs]{Section 7}.

\section{Observations and Data Processing}
\label{sec:obsndata}

\subsection{Data Acquisition}

The \textit{JWST} data presented in this article are obtained from the Mikulski Archive for Space Telescopes (MAST) at the Space Telescope Science Institute. The specific observations analysed can be accessed via \dataset[doi: 10.17909/0js8-gs60]{https://doi.org/10.17909/0js8-gs60}.

As part of GTO Program 1294 (PI: Chen), we use the Mid-Infrared Instrument (MIRI) Medium Resolution Spectrograph (MRS) \citep{Wells15, argyriou2023} to observe $\eta$ Tel A (A0V, K=5.01) \citep{Houk1975, cutri2003} on May 13, 2023. The MIRI MRS is comprised of four IFU channels, with a wavelength-dependent FOV that increases in size per channel; i.e. the FOV is 3.2$\times$3.7" for Channel 1, 4.0$\times$4.8" for Channel 2, 5.2$\times$6.2" for Channel 3, and 6.6$\times$7.7" for Channel 4. Each channel is further divided into three gratings, which cover the short (A), medium (B), and long (C) wavelength ranges of the channel. The total wavelength coverage of the instrument is from 4.9--28 $\mu m$. Since our observation uses all four IFU channels, we also observe $\eta$ Tel B in Channels 3 and 4 due to their larger FOVs. 

To avoid saturation by the primary, we use the FASTR1 readout pattern. 
For the MRSSHORT detector (comprised of Channels 1 and 2), we use 5 groups per integration with 17 total integrations. For the MRSLONG detector (comprised of Channels 3 and 4), we use 17 groups per integration with 6 total integrations. From the \textit{JWST} Exposure Time Calculator (ETC, Pontopiddan et al. 2016), we expect the SNR for $\eta$ Tel A using such an observing set-up to be $\sim$650 at 5.35 $\mu m$ (ETC Workbook 171617). As the MRS is Nyquist-sampled only at the central wavelengths of the detector, we employ a 4-point point-source dither pattern for each exposure to achieve Nyquist sampling across the detector \citep{law2023}; this also mitigates effects from bad pixels and cosmic rays. The resulting total exposure time is 1121.116 s for channels 1 and 2, and 1187.717 s for channels 3 and 4. At the time of observation, the position angle of the aperture used was 283$^{\circ}$.

The recommended observing sequence (following \cite{Worthen2024}) for high-contrast imaging with MIRI MRS is to take a background observation, immediately followed by a science observation, and then a calibration star observation. Such an observing sequence enables background and classical reference PSF subtraction, which are necessary to eliminate background noise and to recover the spatially resolved disk and the brown dwarf companion. It is known that the MIRI MRS receives significant background emission across its wavelength range, with contributions from zodiacal light and the Milky Way dominating at $\lambda<12.5 \mu m$, and contributions from thermal self-emission of the telescope itself dominating at $\lambda>12.5 \mu m$ \citep{Rigby2023}. The behaviour of the thermal background shows a time-dependency that is currently not well-modelled; thus, it is preferable to take background and reference observations close in time to the science observation to perform background subtraction. 

Unfortunately, due to constraints as a GTO program with a fixed amount of telescope time, we took only the science observation for $\eta$ Tel. Since no dedicated background observation was taken, we search MAST for a publicly available background observed as close in time as possible to our $\eta$ Tel observation. We elect to use the SMP-LMC-058 background observation from Program 1532, taken three days before our data on May 10, 2023. The SMP-LMC-058 background consists of a single exposure with 45 groups per 1 integration for each channel, giving a total exposure time of 124.88 s per channel. This is a much larger number of groups per integration than our $\eta$ Tel observation, which has 5 groups for Channels 1 and 2, and 17 groups for Channels 3 and 4.

Likewise, no dedicated PSF reference observation was taken for $\eta$ Tel. To optimise PSF subtraction, it is important to use a reference source that is similar in spectral type and brightness to the science target, so that the PSF is measured with a similar SNR. A PSF reference observation of N Car (A0II, K=4.218) \citep{Houk1975, cutri2003} was already taken for this program several months earlier, to enable PSF subtraction for observations of $\beta$ Pic \citep{Worthen2024}. The N Car observation consists of 4 exposures in a 4-point point-source dither pattern, each with 5 groups per integration for MRSSHORT and 15 groups per integration for MRSLONG. The total exposure time for N Car is 1853.73 s for Channels 1 and 2, and 1764.93 s for Channels 3 and 4. Since N Car is similar in spectral type and magnitude to $\eta$ Tel A (an A0V star with K=5.01), we elect to use the observation of N Car as a reference for PSF subtraction. 

For N Car's background, we use the dedicated $\beta$ Pic background observation taken as part of the same observing sequence in order to maintain contemporaneity. This background observation consists of two exposures in a 2-point dither pattern optimised for extended sources, with the same number of groups per integration as for N Car. This gives a total exposure time of 263.63 s for each of the four channels. Both of these observations were taken on January 11, 2023. A more detailed description of this observing sequence is provided in \cite{Worthen2024}.

Target acquisition is performed for both $\eta$ Tel A and N Car using the stars themselves, so that the target is well centred within the FOV. This is done to minimise the difference between the two pointings, since effects like fringing can be corrected with varying degrees of effectiveness depending on the offset \citep{argyriou2023}. 

\subsection{Data Reduction}

We reduce the raw data using version \texttt{1.14.0} of the \textit{JWST} Spectroscopic Pipeline, with CRDS context \texttt{jwst\_1223.pmap}. We use the same pipeline set-up for the $\eta$ Tel science, N Car reference, and both background observations. The pipeline comprises of three key stages: \texttt{Detector1}, \texttt{Spec2}, and \texttt{Spec3}. The \texttt{Detector1} stage applies detector-level corrections to the raw data for each individual exposure by fitting accumulating counts (`ramps') into count-rates ('slopes'). Since the background estimates are different based on the number of groups per integration, with the threshold being at around 20 groups/integration, it is necessary to ensure at this stage that the number of groups/integration being included is the same between our science and science-background observations (D. Law, private communication). The SMP-LMC-058 background has 45 groups/integration, which is much higher than both the 5 groups/integration for Channels 1 and 2, and the 15 groups/integration for Channels 3 and 4, of our $\eta$ Tel science observation. As such, we customise the \texttt{saturation.py} script from the pipeline so that only the first 5 groups for the MRSSHORT detector (Channels 1 and 2), and the first 15 for MRSLONG (Channels 3 and 4), are used when running \texttt{Detector1} on the raw SMP-LMC-058 background data. We also set the jump detection threshold step from 3-$\sigma$ to 100-$\sigma$ to prevent the introduction of artefacts into the calibrated data, which occurs due to an over-flagging of jumps in the raw data when using the default pipeline settings. 

At the \texttt{Spec2} stage, specific instrument calibrations are applied to the individual exposure outputs from \texttt{Detector1}, in order to calibrate the data into physical astrometric and brightness units. Additionally, for the background observations, a 1D spectrum is extracted for each exposure. We do not make any changes to the default pipeline settings for this stage. 

The \texttt{Spec3} stage takes the corrected exposures from \texttt{Spec2} and combines the 4 dither positions per exposure into a single 3D spectral cube, consisting of one wavelength axis and two spatial axes. We build cubes separately for each of the 12 MIRI MRS sub-bands to avoid averaging different measurements from each of the three wavelength gratings across the four IFU channels. Master background subtraction from the background spectra extracted in \texttt{Spec2} is also applied at this stage. In the \texttt{cube\_build} step, we set the coordinate system to `\texttt{ifualign}' in order to avoid interpolation of the cubes from the instrument to sky frames, as well as to facilitate subsequent PSF subtraction using the science and reference cubes. We build our spectral cubes using the \texttt{drizzle} algorithm \citep{law2023}, retaining the default pipeline pixel sizes for each channel (i.e. 0.13" for Channel 1, 0.17" for Channel 2, 0.20" for Channel 3, and 0.35" for Channel 4).

\begin{figure}[ht]
    \centering
    \includegraphics[width=0.45\textwidth]{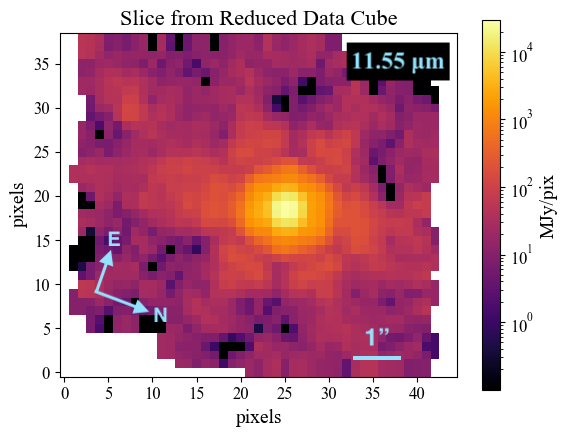}
    \caption{An example slice from the calibrated sub-band 3A data cube output by Stage 3 of the \textit{JWST} pipeline, shown with a logarithmic scaling. Note that the six-point PSF dominates most of the FOV. Blacked-out NaN values within the FOV indicate areas of slight over-subtraction due to pipeline background subtraction.}
    \label{fig:1}
\end{figure}

\begin{figure}[ht]
    \centering
    \includegraphics[width=0.45\textwidth]{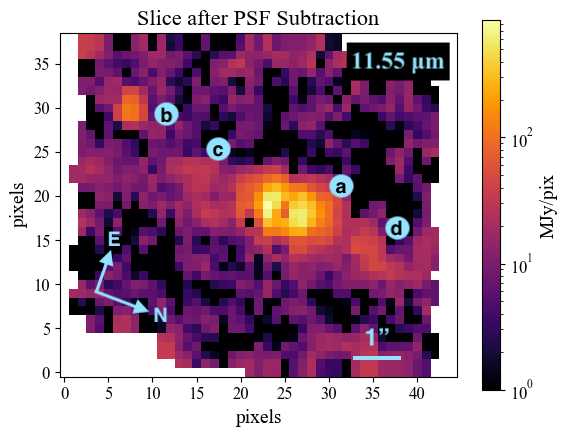}
    \caption{The same wavelength slice shown in \autoref{fig:1} after PSF subtraction using N Car, in which the PSF has been scaled to the peak flux of the $\eta$ Tel slice. The image is log-scaled, so the blacked-out values within the FOV indicate regions of slight over-subtraction. Astronomical objects are labelled: (a) the spatially resolved component of the $\eta$ Tel A debris disk, (b) $\eta$ Tel B, (c) background galaxy 2CX0 J192251.5-542530, and (d) an unknown feature (see \hyperref[ssec:Adiskspat]{Section 3.3}).}
    \label{fig:2}
\end{figure}

\begin{figure*}[ht]
    \centering
    \includegraphics[width=\textwidth]{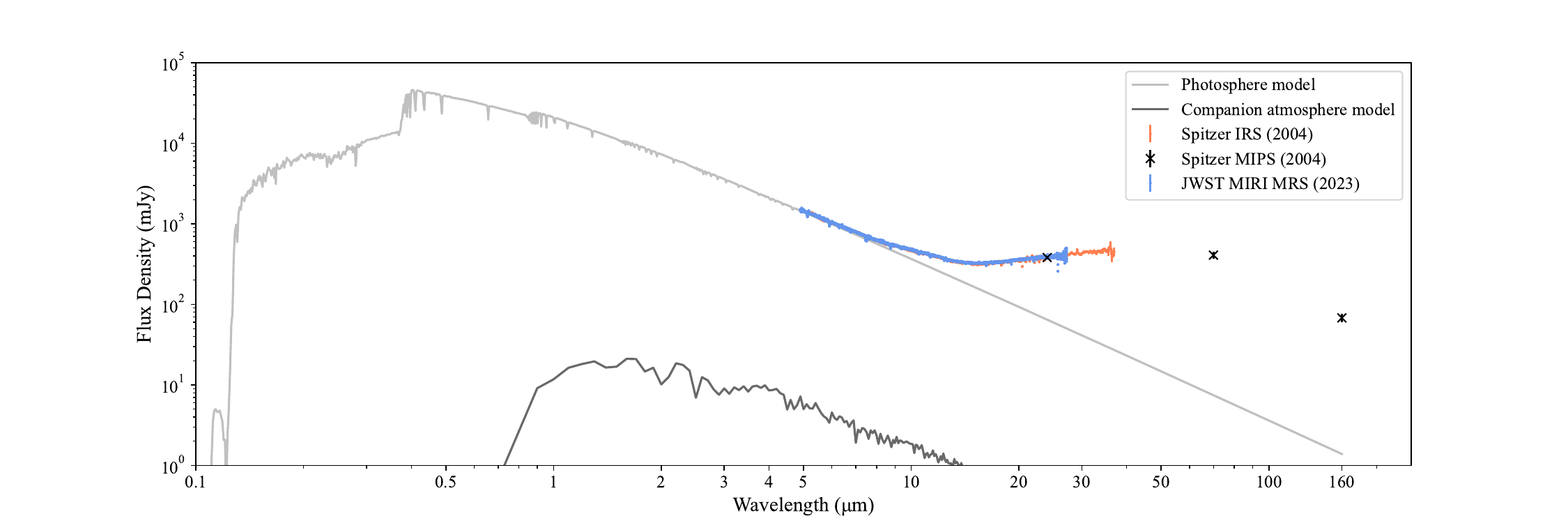}
    \caption{MIRI MRS spectrum of $\eta$ Tel A, overplotted onto a $T_\mathrm{eff}=9700$ K, $log(g)=4.0$ photosphere model \citep{Mittal2015}. The \textit{Spitzer} IRS spectrum \citep{Chen14} is shown in orange, along with MIPS photometry \citep{Rebull2008}. We include our atmosphere model for $\eta$ Tel B (see \hyperref[ssec:astrometry]{Section 5.1}) to show that flux contribution from the companion does not significantly impact the spectrum of the disk even with the larger \textit{Spitzer} aperture.}
    \label{fig:3}
\end{figure*}

\subsection{PSF Subtraction}

The resolved disk and the companion are both $\sim$$10^{-4}$ times of magnitude fainter than the primary star \citep{Smith2009, Lowrance2000}. Thus, to recover the spatial extent of the debris disk around $\eta$ Tel A, as well as improve the S/N at which the brown dwarf companion is detected, we perform classical reference PSF subtraction on the calibrated data cubes output by the pipeline. To do this, we calculate the centroids for both $\eta$ Tel A and N Car by fitting a 2D Gaussian to each wavelength slice in the cubes. Averaging over the centroids for all slices in the cube gives us the final centroid positions for each cube. 
We then interpolate the N Car cubes to the nearest value so that the location of the N Car centroid in each slice aligns with that of the $\eta$ Tel centroid. After scaling the flux of the N Car slices to an $\eta$ Tel A photosphere model from \cite{Mittal2015}, we finally perform a slice-by-slice subtraction of the N Car cube from the $\eta$ Tel cube. We scale the PSF to the $\eta$ Tel photosphere to obtain the total flux contribution of the disk; however, in using this scaling, we do not see the double-lobed structure reported in \cite{Smith2009}. We then apply a second PSF scaling, following the method used by \cite{Smith2009}, in which we scale the flux of the N Car slices to the peak flux of the observed $\eta$ Tel slices, before subtracting N Car from $\eta$ Tel. With this scaling, we are able to recover the spatially resolved flux component of the debris disk, and identify the two lobed structure expected from a compact, edge-on disk. 

\hyperref[fig:1]{Figures 1} and \hyperref[fig:2]{2} show an example slice of the calibrated MIRI MRS $\eta$ Tel data before and after the peak-flux scaled PSF subtraction, with the latter indicating the location of all astronomical objects within the FOV. In both scaling methods, the brown dwarf companion can be seen in the top-left corner of the IFU-aligned cubes across sub-bands 3A to 4A ($\sim$11--21 $\mu m$). Additionally, using the second scaling method, we detect the presence of background galaxy 2CX0 J192251.5-542530 within the FOV in sub-bands 1A to 3C, and a second, previously undetected, extended source that is likely a background galaxy in sub-bands 1A to 3A.

The calibrated, PSF-subtracted spectral cubes are the final data products which we use for our following analysis. 

\section{The $\eta$ Tel A Debris Disk}
\label{sec:Adisk}

\subsection{A New Epoch of Mid-Infrared Spectroscopy}
\label{ssec:Aspec}

We extract the $\eta$ Tel A spectrum over 5--29 $\mu m$ by performing point-source aperture photometry with the \texttt{\text{spec3.extract\_1d()}} function of the \texttt{jwst} pipeline. We set our aperture radius to be 2.0 $\times$ FWHM. As the pipeline-produced spectrum shows slight vertical offsets between the MRS sub-bands, particularly at longer wavelengths (i.e. Channels 3 and 4), we also perform absolute flux calibration by applying a Relative Spectral Response Function (RSRF), as described by:

\begin{equation}
    \textrm{RSRF}=\frac{\textrm{reference~model~spectrum}}{\textrm{reference~extracted~spectrum}}
    \label{eq:1}
\end{equation}

For the RSRF, we use N Car to calibrate our observations. We use an N Car photosphere model from \cite{Worthen2024} as the model spectrum, and extract a spectrum from the MRS observations of N Car across Channels 1--4 using the same aperture size as for our $\eta$ Tel A extraction. We truncate the calibrated spectrum at 25.63 $\mu m$, beyond which noise significantly worsens the SNR.

Our calibrated 5--26 $\mu m$ MIRI MRS spectrum for $\eta$ Tel A is shown in \autoref{fig:3}.
We find that the MIRI MRS spectrum is consistent with the updated reduction of the 2004 \textit{Spitzer} IRS spectrum \citep{Chen14}, indicating that the disk has not noticeably evolved over time.
\begin{figure}[h!]
    \centering
    \vspace{-2mm}
    \includegraphics[width=0.45\textwidth]{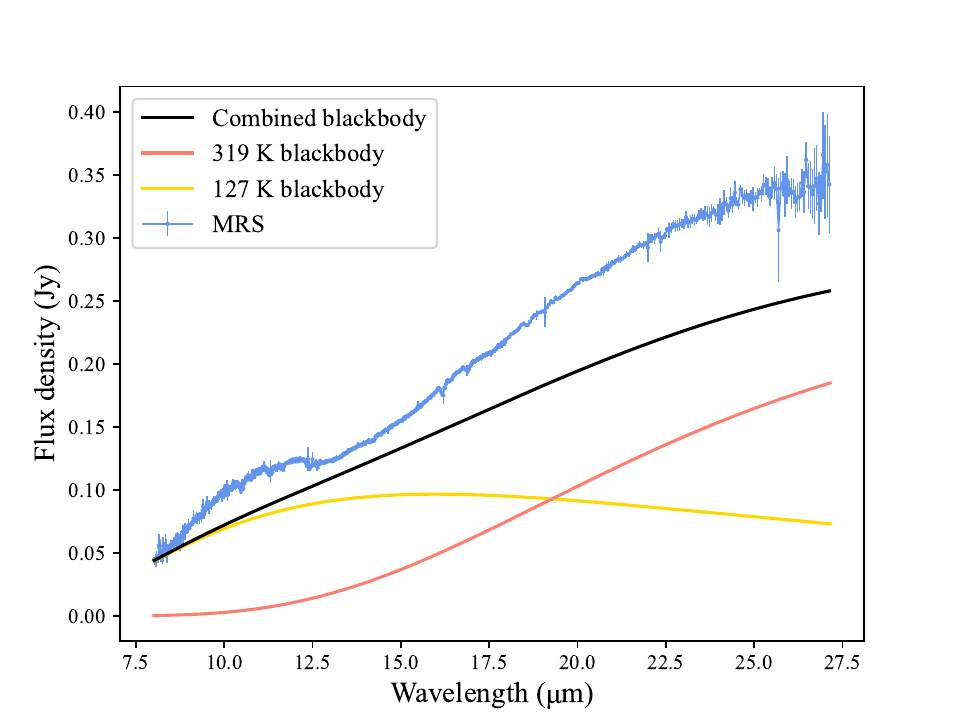}
    \caption{Photosphere subtracted spectrum of $\eta$ Tel A, clearly showing the 10 $\mu m$ feature weakly detected by \textit{Spitzer} IRS, as well as a broad 20 $\mu m$ detected for the first time in this disk. The best-fit model for total contribution from the continuum is shown in black; the continuum is best fit by two blackbody components corresponding to a warm dust population with $T_w=319\pm59$ K and a cool population with $T_c=127 \pm 25$ K. This two-component structure is consistent with previous modelling of the disk \citep{Chen2006, Smith2009}.}
    \label{fig:4}
\end{figure}
\begin{figure*}[ht]
    \centering
    \includegraphics[width=\textwidth]{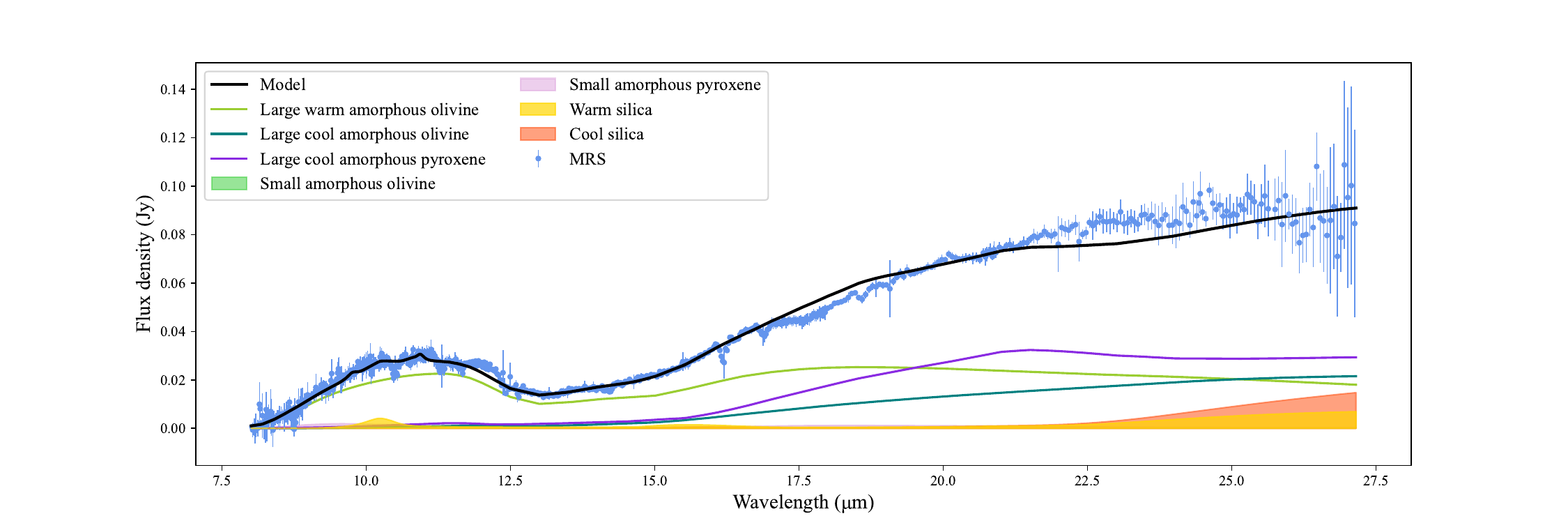}
    \caption{Photosphere $+$ continuum subtracted spectrum of the $\eta$ Tel A disk, binned by a factor of 10. There is a firm detection of the 10 and 20 $\mu m$ dust component features. Modelling indicates the presence of large amorphous olivine and pyroxene grains, as well as some silica, in the disk. The model's difficulty with fitting the tail end of the broad 20 $\mu m$ feature may be due to the presence of additional components that we do not account for; more detailed modelling outside the scope of this work may help to identify these components.}
    \label{fig:5}
\end{figure*}

Although the lower angular resolution of the IRS means that the flux of the brown dwarf companion is included in its aperture, we do not consider the IRS spectrum of $\eta$ Tel A to be significantly impacted by flux from $\eta$ Tel B since the primary is at minimum $\sim$$10^3$ brighter than the companion over the IRS wavelength range, as indicated by comparison with our atmosphere model for the brown dwarf (see \hyperref[ssec:Batmomod]{Section 4.1} for modelling details).

Following photosphere subtraction from the MIRI MRS spectrum, we clearly recover the 10 $\mu m$ silicate feature suggested by \cite{Chen2006}, and identify a broad 20 $\mu m$ feature for the first time in this disk. Broad spectral features at 10 and 20 $\mu m$ have been observed in many debris disks and T Tauri stars, with the 20 $\mu m $ feature being fairly common in cases where a 10 $\mu m$ feature is present (e.g. \cite{Chen2006}, \cite{sargent2009ApJ}). These two features are indicative of the presence of amorphous silicates, which are known to show broad spectral bands at 10 $\mu m$ due to Si-O stretching and 20 $\mu m$ due to O-Si-O vibrations \citep{Henning2010}. Indeed, previous modelling of the \textit{Spitzer} IRS data predicted a 20 $\mu m $ dust component contribution to the overall spectrum for a composition of large amorphous olivine grains \citep{Chen2006}, although no such feature was evident in the IRS data, likely due to its lower SNR.

\subsection{Dust Modelling}
\label{ssec:Adustmod}

To better understand the disk's spectral features, we perform detailed modelling of the new MIRI MRS spectrum over 7.5--26.9 $\mu m$ using code originally developed by \cite{sargent2009ApJ} to model silicate and silica features in the \textit{Spitzer} IRS spectra of T Tauri stars. We truncate the MRS spectrum shortwards of 7.5 $\mu m$, as the fitting code attempts to reproduce wiggles in the spectrum at shorter wavelengths created by the incomplete correction of stellar absorption features. We also truncate the spectrum longwards of 26.9 $\mu m$ as the spectrum becomes substantially noisier at longer wavelengths.

The dusty disk around $\eta$ Tel is believed to contain dust grains that radiate as a featureless continuum \citep{Chen2006}. We assume a simplified case of two `bands' of dust populations, and set uniform priors on the temperatures of the cold and warm dust populations to $T_c=80$--200 K and $T_w=201$--800 K respectively. These temperature ranges are divided into 7 steps (i.e. 17 K and 86K increments), which are explored over to determine the best fit. We obtain a best-fit warm black body dust temperature of $T_w = 319\pm59$ K and a cool black body dust temperature of $T_c = 127 \pm 25$ K; here, the uncertainties do not represent the 1-$\sigma$ confidence level, but rather the temperature fitting precision given the prior range on dust grain temperatures and the number of temperature bins used for the fit. \autoref{fig:4} shows the warm and cool black body components overlaid onto our photosphere subtracted MRS spectrum to show the contribution of the black body continuum to the overall spectrum. We find that the best-fit warm and cool dust temperatures are broadly consistent with those found for the 2004 \textit{Spitzer} IRS spectrum ($T_w = 370$ K, $T_c =115$ K; \cite{Chen2006}).

To more clearly view the spectral features, we then subtract the continuum. Previous modelling of the \textit{Spitzer} IRS $\eta$ Tel spectrum suggested that the 10 $\mu m$ emission was due to the presence of amorphous silicates in the disk's warm dust population \citep{Chen2006}. Modelling also suggested that the same warm amorphous silicates could give rise to emission at 20 $\mu m$, although such a feature was not detected in the \textit{Spitzer} IRS data.

We model the 10 and 20 $\mu m$ spectral features and find that large warm amorphous olivine is the primary contributor to the 10 $\mu m$ feature, consistent with the literature. The broad 20 $\mu m$ feature, however, appears to be best fit by primarily a combination of large warm and cool amorphous olivine, large cool amorphous pyroxene, with some contribution from warm and cool silica at longer wavelengths. Our best-fit model for the 10 and 20 $\mu m$ spectral features is shown in \autoref{fig:5}. The $\chi^2_\mathrm{red}$ for the entire spectral fit is $3.8$. We note that the 20 $\mu m$ feature appears shifted to a slightly longer wavelength compared to the model. This, along with the presence of some small peaky structures in the 10 $\mu m$ feature, suggests that there may be additional dust components contributing to the data that we have not presently accounted for in our modelling. More detailed modelling, which is outside the scope of this work, may help to resolve this mismatch between the model and the data.

\subsection{The Spatial Distribution of Dust in the Disk}
\label{ssec:Adiskspat}

\begin{figure*}
    \centering
    \includegraphics[width=0.9\textwidth]{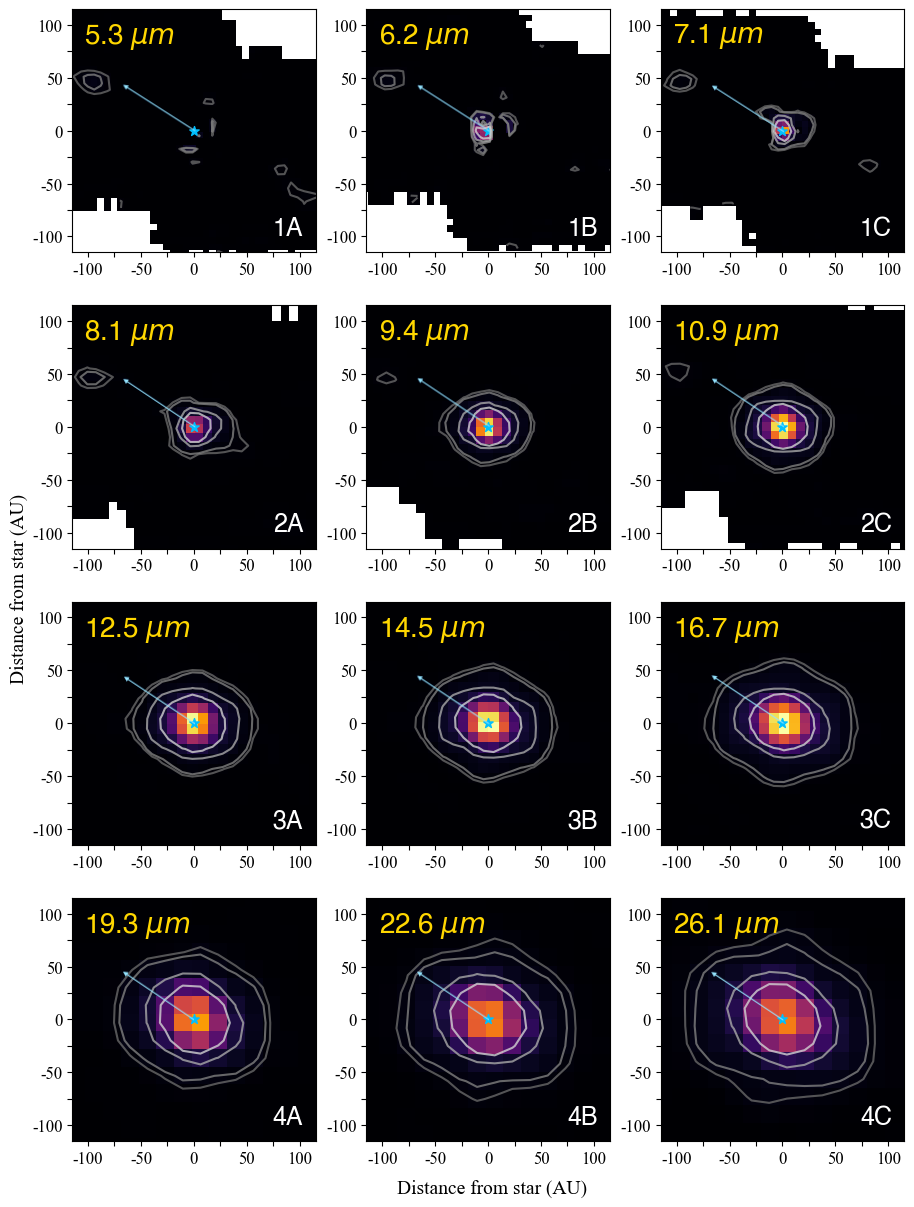}
    \caption{PSF-subtracted, collapsed-cube images of $\eta$ Tel, in which the PSF was scaled to the photosphere of the star to reveal the total flux contribution from the disk. Images are linearly scaled from $0$--$10^4$ MJy/sr and zoomed in on the disk. The contour lines indicate regions at 3-, 5-, 20-, 50-$\sigma$ detection thresholds for each sub-band image. The location of the primary is marked with a blue star, and the white arrow indicates the direction of the brown dwarf companion. The wavelength labels give the central wavelengths of each sub-band to which each image has been collapsed. Note that the radial extent of the disk appears to increase with wavelength.}
    \label{fig:6}
\end{figure*}
\begin{figure*}
    \centering
    \includegraphics[width=0.9\textwidth]{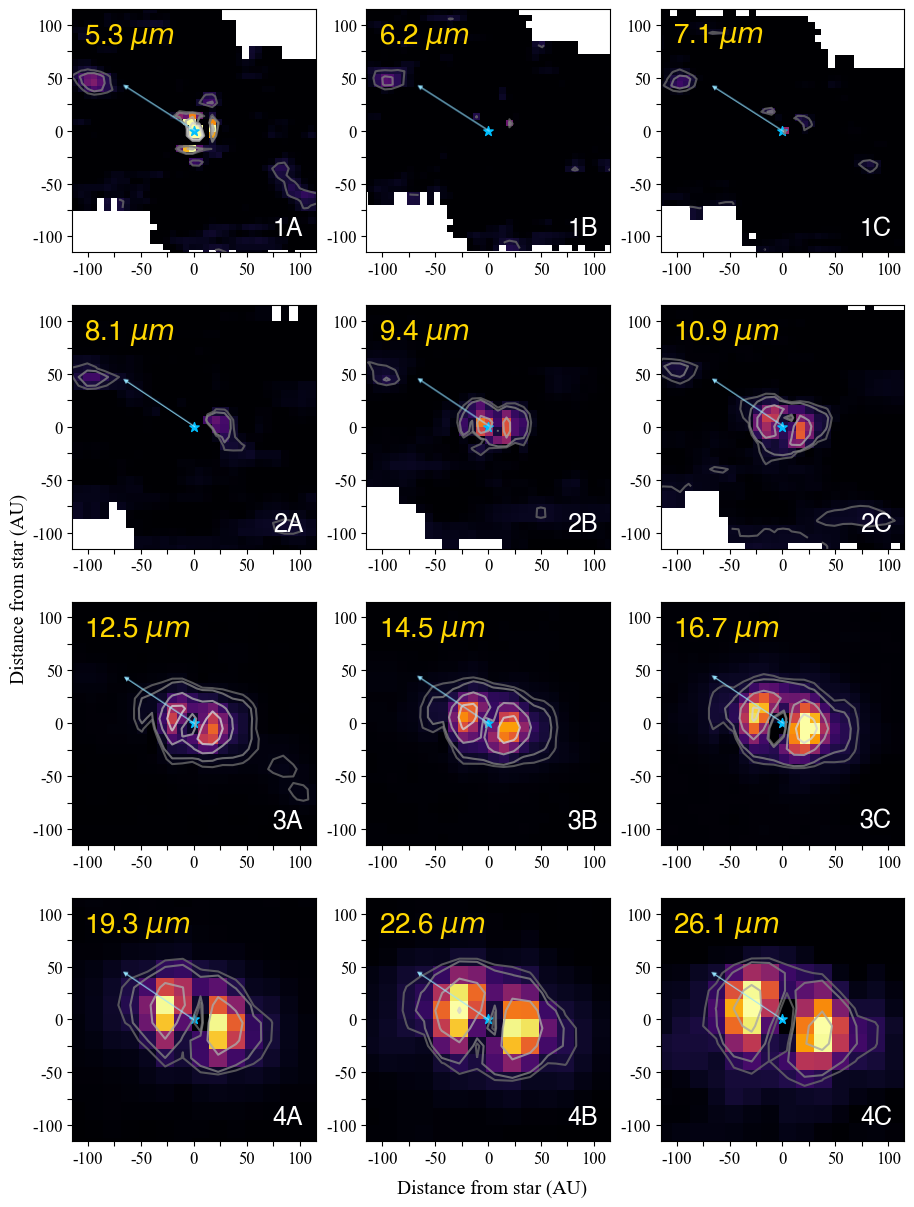}
    \caption{Similar to \autoref{fig:6}, except the PSF has been scaled to the peak flux of the science image before being subtracted in order to reveal the spatially resolved dust component. The images are displayed with a linear scaling from $0$--$1.15\times10^3$ MJy/sr. We observe that the disk appears largely axisymmetric across all wavelengths, and that there appears to be a slight offset in PA between the disk and the position of $\eta$ Tel B. We also note the apparent increase in radial extent and inner cavity size of the disk with wavelength. The latter only appears in this scaling and is likely an artefact introduced by PSF subtraction.}
    \label{fig:7}
\end{figure*}

Following PSF subtraction of the science images, we are able to obtain information about the spatial distribution of dust in the disk from 8.67--27.89 $\mu m$ (MRS sub-bands 2B to 4C). By scaling the PSF to the photosphere of the star before subtraction, we can recover the total flux contribution from the dust in the disk. In these images, unresolved excess flux dominates. As a result, we perform a different scaling of the PSF to the peak flux of the science images before subtraction, as in \cite{Smith2009}, to highlight the spatially extended emission. \autoref{fig:6} and \hyperref[fig:7]{7} show the collapsed cube images of the disk resulting from these two PSF scalings. We sum up all the wavelength slices in each cube in order to obtain the highest possible SNR. 

We assess which features in the cubes are due to real disk morphology, and which are due to artefacts. At short wavelengths, we find inconsistent structures close to the location of the star in both scalings; these are likely due to PSF residuals. The ellipsoidal feature in the upper left corner of the Channel 1 \& 2 sub-bands is background galaxy 2CX0 J192251.5-542530. We note the 3-$\sigma$ detection of two small unknown features: one to the immediate right of the primary across across sub-bands 1A--2A, and one in the lower right corner of 1A--1C and 3A. The detection of these features across several sub-bands makes them unlikely to be due to warm pixels. However, interpreting their spectra has proved challenging due to their low S/N and significant discontinuities between sub-bands; as such, the sources of the two features remain inconclusive.

Emission from the disk itself starts to pick up from $\sim$9 $\mu m$ (sub-band 2B) onwards. We observe an apparent increase in radial extent of the disk with wavelength, which has two potential explanations. If real, this could be due to the increased sensitivity at longer wavelengths to cooler dust populations farther out from the star. This may indicate that the $\eta$ Tel disk possesses a more continuous structure, contrary to the 2-component structure suggested by \cite{Smith2009}, which consists of a narrow ring of material at a fixed distance of $\sim$24 au from the star, along with an unresolved flux component at inwards of $\sim$4 au. Alternatively, the apparent radial increase could be an artefact introduced by the increase in pixel and PSF size with wavelength `smearing' the flux from the dust, thus causing it to appear farther out from the star at longer wavelengths. 

As resolving this degeneracy will require modelling that is beyond the scope of this work, we focus our considerations on the case for the disk with a 2-component structure, as held in the literature \citep{Chen2006, Smith2009}. 

In the second scaling, we also note an apparent increase in size of an inner cavity between the lobes with wavelength. Since this apparent increase is not observed in both scalings, we conclude it is likely an artefact introduced by the PSF subtraction, due to the increasing pixel and PSF size with wavelength. 

\section{Does $\eta$ Tel B have an infrared excess?}
\label{sec:Bdisk}

Since $\eta$ Tel B is young, and both $\eta$ Tel A and HD 181327 are known to host debris disks, it is natural to wonder if $\eta$ Tel B likewise possesses a debris disk. Motivated by this question, we seek to determine if the companion possesses an infrared excess at longer wavelengths indicative of the presence of warmed circumstellar dust.
 
\subsection{$\eta$ Tel B Atmosphere Modelling}
\label{ssec:Batmomod}

To understand the extracted spectrum of $\eta$ Tel B from the MIRI MRS data, we must first understand the companion's expected atmosphere. We model the atmosphere of $\eta$ Tel B with the \texttt{species} package \citep{Stolker2020} by fitting existing spectra and photometry
for the brown dwarf companion to a BT-SETTL (CIFIST) model grid \citep{allard2011}. Our model fit uses spectroscopic measurements from HST/STIS \citep{Lowrance2000} and VLT/SINFONI \citep{Bonnefoy2014}, as well as photometric measurements from HST/NICMOS $H$ band \citep{Lowrance2000} and Gaia $G$ band \citep{GAIA2023}. Additionally, we use photometry derived from the observed magnitude difference of $\eta$ Tel A and B in the following instrument filters: HST/NICMOS F110W \citep{Neuhauser2011}, Paranal/ISAAC $K$ band \citep{Guenther2001}, Paranal/NACO $H$, $K$, and $L_p$ bands \citep{Neuhauser2011}, and VLT/VISIR \textit{PAH} \citep{Geissler2008}.  

We vary parameters for $T_{\mathrm{eff,}B}$, $log(g)$, radius, and parallax, setting uniform priors for the first three (\autoref{tab:1}), and Gaussian priors on the parallax and the companion's mass. We assume that the parallax is the same as that for the primary, given as 20.6028$\pm$0.09 mas in Gaia EDR3 \citep{GAIA2023}. We set a companion mass prior of $M_B=$ 35$\pm$15 $M_{Jup}$, following \cite{Neuhauser2011}.

\begin{figure*}[ht]
    \centering
    \includegraphics[width=\textwidth]{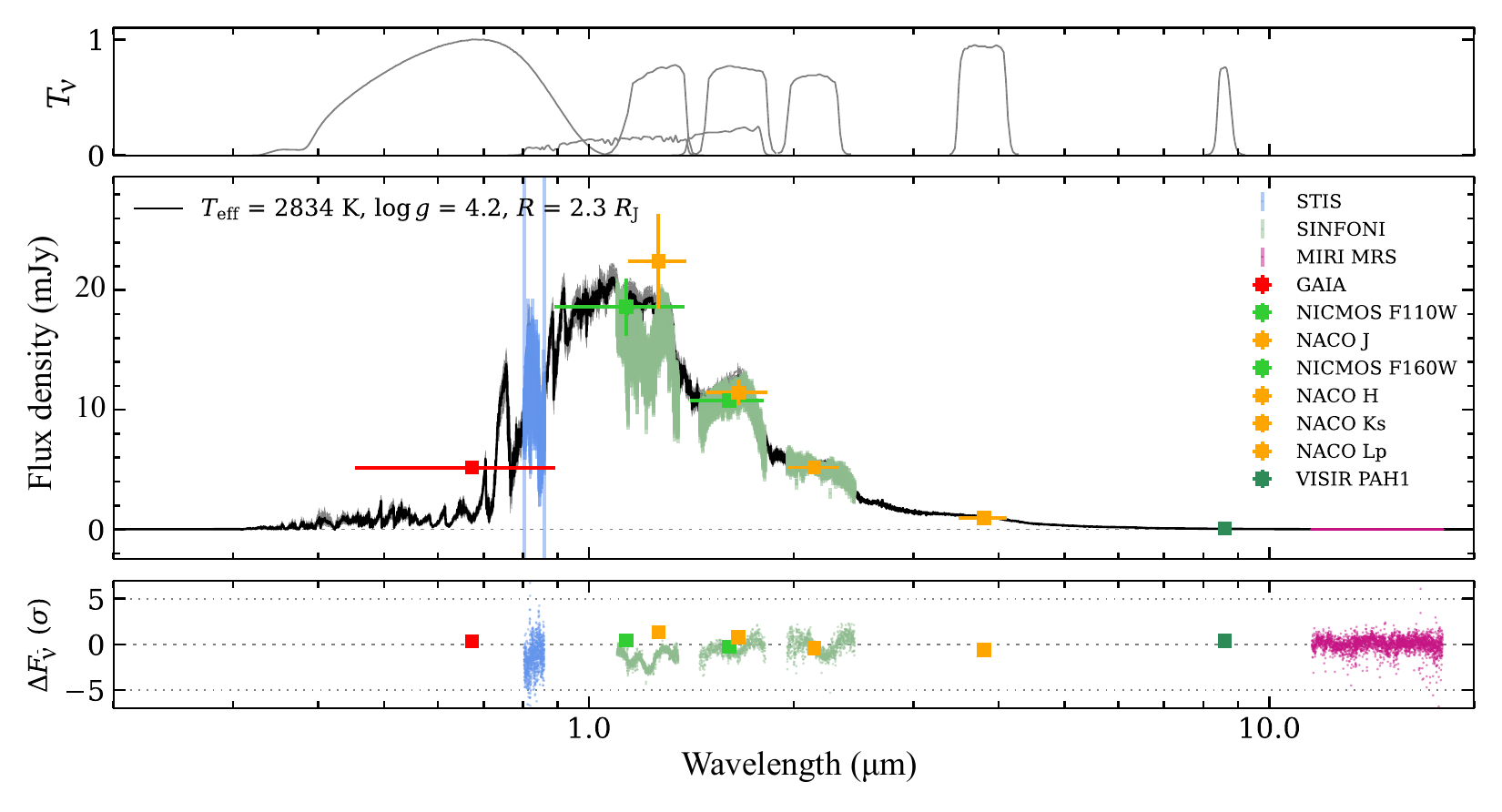}
    \caption{(\textit{Top panel}) Transmission profiles for each photometric filter, indicating flux density throughput versus bandwidth on a normalised scale, where 1 is full flux transmitted and 0 is no flux transmitted. (\textit{Centre panel}) The full 0.85--21 $\mu m$ SED of $\eta$ Tel B, showing existing spectra and photometry along with the new MIRI MRS 11--20 $\mu m$ spectrum. Black line shows the best-fit atmosphere model for the companion, which was calculated by fitting the spectroscopic and photometric data to a BT-SETTL (CIFIST) model grid using \texttt{species} \citep{Stolker2020}. $T_\mathrm{eff}$, $\log{g}$, and $R$ values are consistent with expectations for a M7/8-type brown dwarf (see \autoref{tab:1}). (\textit{Bottom panel}) Residual flux density for each data point from the literature compared to the best-fit model in $\sigma$. All residuals are within $\pm$5-$\sigma$, indicating that the model is a good fit to the data.}
    \label{fig:8}
\end{figure*}

The data is weighted such that each dataset, spectroscopic and photometric, is equal. This prevents each point in the spectroscopic datasets from being weighted equally to each photometric point. We use the nested sampling algorithm \texttt{UltraNest} \citep{Buchner2021} to sample 300 live points from the prior. \autoref{tab:1} summarises the priors used in our model fitting, and the best-fit parameters for $\eta$ Tel B.
The resulting atmospheric model, calculated at the native resolution of MIRI MRS ($R\sim2700$) is shown in \autoref{fig:8} at $R=1000$. 

We note that our derived companion mass of $M_B=29^{+16}_{-13} M_J$ is lower than the $M_B=47^{+5}_{-6} M_J$ value obtained by \cite{Lazzoni2020} using AMES-COND models \citep{Baraffe2003}, although it is consistent when considering both sets of error bars. This discrepancy in mass may be due to the fact that we do not account for the age of the system in our atmospheric modelling.

\begin{table}
    \begin{center}
        \caption{Priors and Best-fit Parameters for $\eta$ Tel B}
        \begin{tabular}{ccc}
            \hline
            Model Parameter&  Prior Range&  Best-Fit\\
            \hline
            $T_{\mathrm{eff,}B}$ [K]&  (2500, 3000)&  2830$^{+20}_{-30}$\\
            $log(g)$&  (3.5, 5.0)&  4.3$^{+0.1}_{-0.2}$\\
            $R_B$ [$R_{J}$]&  (0.5, 5.0) &  2.28$\pm0.03$\\
            $\pi$ [mas]&  20.6028 $\pm$ 0.09&  N/A\\
            $M_B$ [$M_{J}$]& $35\pm15$&  $29^{+16}_{-13}$\\
            $log L_B/L_{\odot}$&  N/A&  $-$2.48$\pm0.01$\\
            \hline
            $\chi^2_{red}$ &  &  5.49\\
            \hline
        \end{tabular}
    \end{center}
    \vspace{-2mm}
    \footnotesize{Note: prior ranges in parentheses indicate uniform distributions, and prior ranges for parallax $\pi$ and companion mass $M_B$ are Gaussian distributions.}
    \label{tab:1}
\end{table}

\begin{figure*}[ht]
    \centering
    \includegraphics[width=0.9\textwidth]{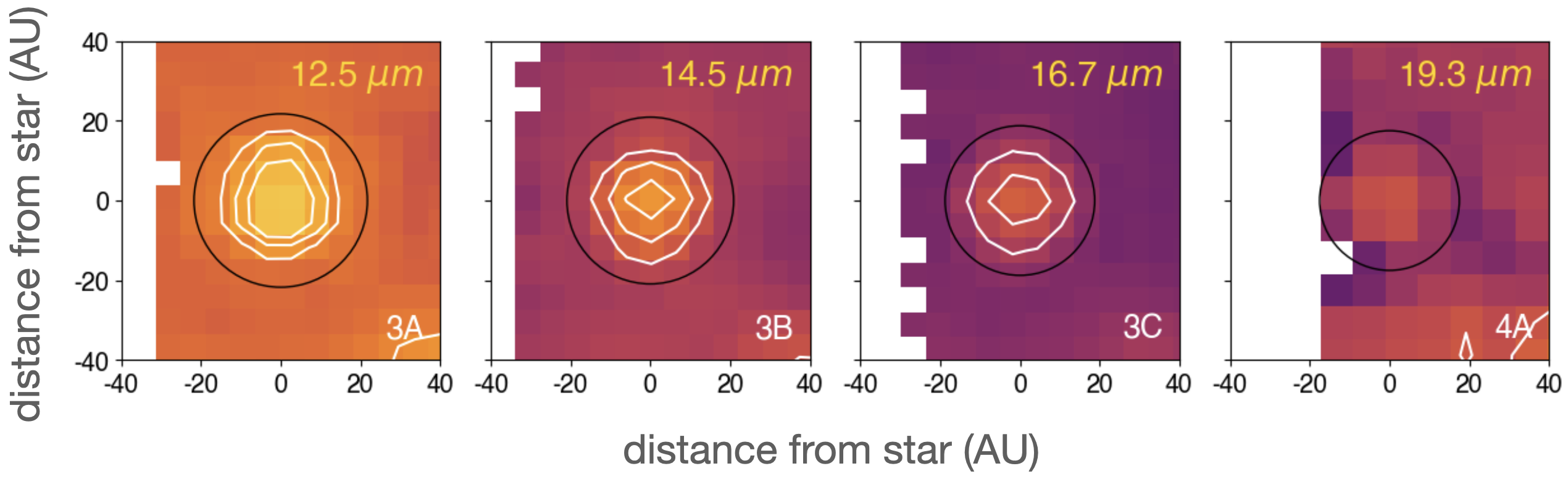}
    \caption{MIRI MRS collapsed sub-band images of $\eta$ Tel B from 3A to 4C using a logarithmic scaling. We obtain a 3--10-$\sigma$ detection of the brown dwarf across 3A to 3C, and a 3-$\sigma$ detection in 4A. In sub-bands 4B and 4C, increased background noise renders the companion irrecoverable; for this reason, we omit these two sub-bands in our spectral extraction of the brown dwarf. The black circle indicates the size of the extraction aperture.}
    \label{fig:9}
\end{figure*}

\subsection{$\eta$ Tel B Spectrum}
\label{sec:bspec}

The detection of warm circumstellar dust around brown dwarfs is dependent on the mass and temperature of the dust, with submillimetre and millimetre observations being most suited to identifying the presence of traditional 100 au sized debris disks \citep{Apai2013}. However, if $\eta$ Tel B does possess a compact debris disk, we may also be able to detect it via an infrared excess across the MRS wavelength range. 

To extract an MRS spectrum of $\eta$ Tel B, we perform aperture photometry at each wavelength slice of the cubes for sub-bands 3A, 3B, 3C, and 4A. We omit sub-bands 4B and 4C, as the increase in background noise and lower instrument throughput renders the companion irrecoverable at these longer wavelengths (\autoref{fig:9}).
Since $\eta$ Tel B is a faint source located at the edge of the MRS FOV, the pipeline does not do a satisfactory job of extracting the spectrum. As such, we manually employ a tapered-column extraction technique for the $\eta$ Tel B unresolved point source. This involves increasing the aperture size over wavelength to account for the diffraction limit being proportional to wavelength ($\theta=1.22\frac{\lambda}{d}$). Due to the relative faintness of $\eta$ Tel B, it is difficult to empirically obtain its FWHM. As such, we use the FWHM calculated for the reference star N Car, since the FWHM of the instrument should behave similarly irrespective of observing target.

\begin{figure}[ht]
    \centering
    \includegraphics[width=0.45\textwidth]{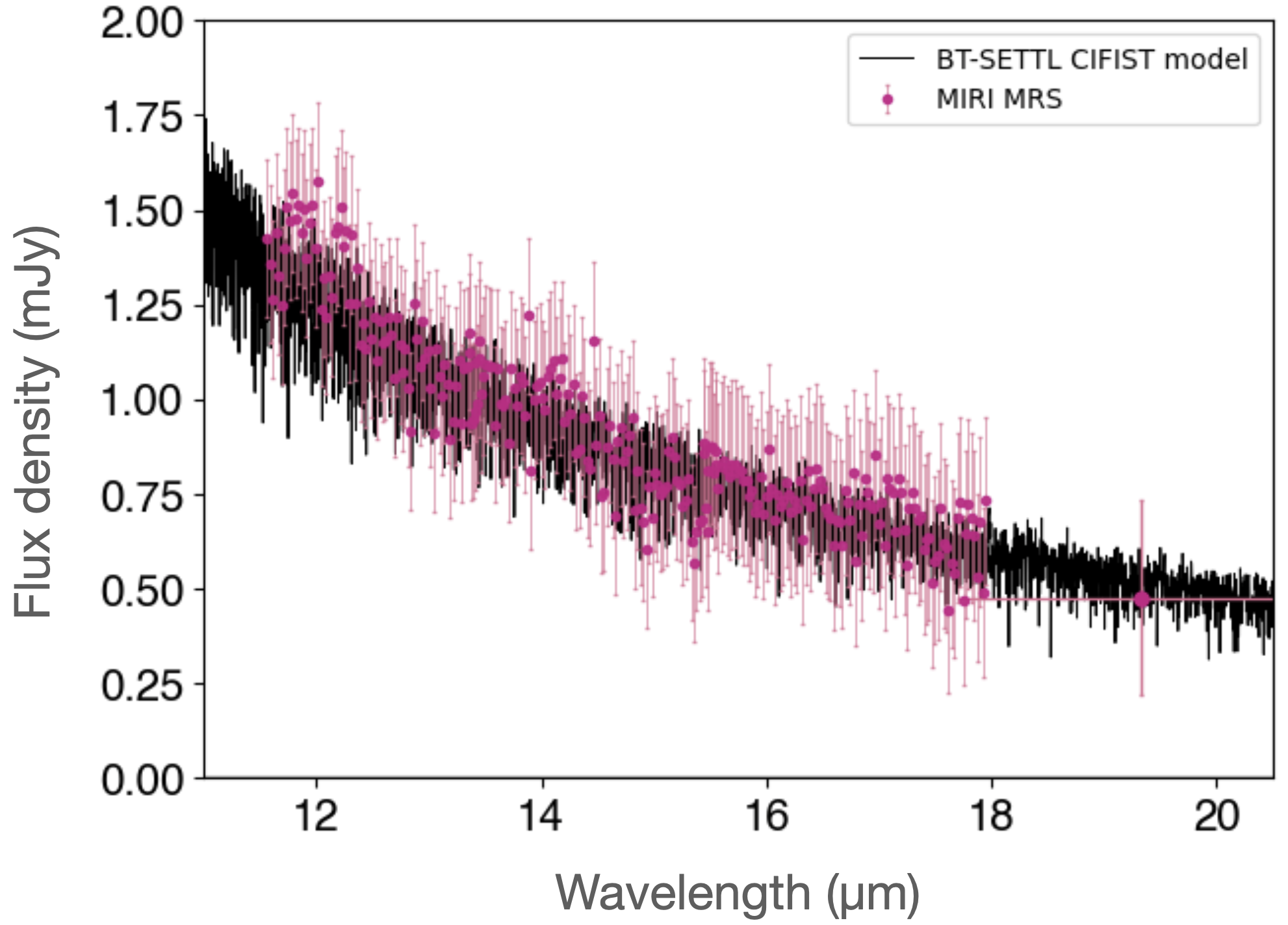}
    \caption{MIRI MRS spectrum of $\eta$ Tel B, compared to the atmosphere model from \hyperref[ssec:Batmomod]{Section 4.1}. The extracted spectrum is consistent with the model, showing no infrared excess.}
    \label{fig:10}
\end{figure}

Additionally, if the aperture is too large, it could include additional noise into our extraction. Thus we restrict the radius of our aperture to be 0.87 $\times$ FWHM in channel 3 and 0.30 $\times$ FWHM in channel 4 (or $\sim$1" on-sky for both channels) multiplied by a factor of $\lambda/\lambda_0$, in order to reduce flux contribution from the background and to improve the SNR of the extraction. We again perform absolute flux calibration and align discontinuities between subbands in the spectra by applying an N Car RSRF; in this case, however, we extract our N Car spectrum across sub-bands 3A to 4A using the same aperture sizes and tapered-column method as for our $\eta$ Tel B extraction. As the calibrated spectrum remains fairly noisy, particularly for 4A, we bin the spectra for sub-bands 3A--3C by a factor of 10, and collapse 4A into a single photometric point. We present the final $\eta$ Tel B spectrum in \autoref{fig:10}, overplotted onto our atmosphere model (see \hyperref[ssec:Batmomod]{Section 4.1}). To calculate the error-bars, we perform an injection-recovery test of N Car; this involves scaling N Car to the model flux of $\eta$ Tel B and injecting it into the $\eta$ Tel Stage 3 cubes on the opposite side of the primary to $\eta$ Tel B, before applying PSF subtraction and extracting its spectra using the exact same methods detailed above. Taking the average residuals between the injected and recovered spectrum for each sub-band gives us the error-bars for that sub-band.

Our spectrum shows a good fit to the atmosphere model. This indicates that $\eta$ Tel B does not possess an infrared excess between 11--24 $\mu m$. We note that this does not necessarily rule out the presence of circumstellar dust around $\eta$ Tel B; the companion's low luminosity may mean that there is very little dust at 100--260 K temperatures, which would make any excess in the 11--24 $\mu m$ range simply too faint to be identified by MIRI MRS. 
However, from our current observations, we conclude that we do not identify the presence of a debris disk around $\eta$ Tel B.

\section{The Orbit of $\eta$ Tel B}
\label{sec:Borb}

\subsection{A New Epoch of Astrometry}
\label{ssec:astrometry}

\begin{figure*}[ht]
    \centering
    \includegraphics[width=\textwidth]{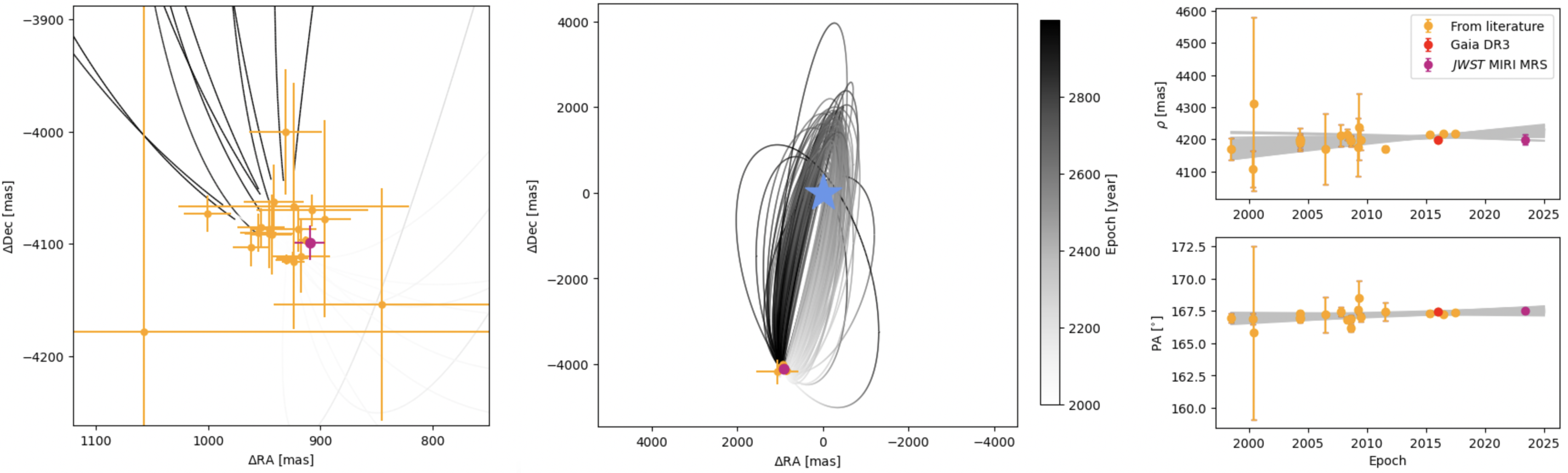}
    \caption{Projected orbits of $\eta$ Tel B from orbital derivation. (\textit{Centre}) On-sky depiction of a sample of 100 potential orbits. The blue star marks the location of the primary and the maroon point marks the current location of the companion from MRS astrometry. (\textit{Left}) Close-up of relative astrometry points clustered in the centre image. Orange points indicate literature values. (\textit{Right}) Separation and position angle versus epoch for the companion with respect to the primary. Due to the long-period nature of the orbit, it is likely that any significant change in separation/PA will not be observable for another few decades.}
    \label{fig:11}
\end{figure*}

The high angular resolution of MIRI MRS allows us to obtain positional accuracy for $\eta$ Tel A and B to 10 and 23 mas for Channels 3 and 4 respectively \citep{Patapis2024}. This enables us to derive a new epoch of relative astrometry, extending the baseline of astrometric measurements by 6 years since the most recent measurement with VLT SPHERE \citep{nogueira2024astrometric}, and by 25 years since the first measurement with \textit{HST} NICMOS \citep{Lowrance2000}.

To do this, we fit a 2D Gaussian to the collapsed image of each sub-band cube from 3A--4A in order to first identify the pixel coordinates of the $\eta$ Tel A and B centroids; we exclude the cubes for sub-bands 4B and 4C due to higher noise levels increasing the uncertainty in the precise location of the $\eta$ Tel B centroid. Transforming the pixel coordinates to right ascension (RA) and declination (Dec) values using the World Coordinate System (WCS) then allows us to calculate the separation and position angle of $\eta$ Tel B with respect to the primary star. 
To obtain the final angular separation and positional angle, we average over the results from sub-bands 3A--4A, estimating the uncertainties as the standard deviation between the measurements. We calculate a final separation of 4199$\pm$15 mas ($\sim$200 au) and a position angle of 167.49$\pm$0.18$^{\circ}$. As both the primary and companion are observed in Gaia DR3 \citep{GAIA2023}, we also use it calculate the Gaia relative astrometry: $\rho=4197.3\pm3.7$ mas and $\theta=167.44\pm0.09^{\circ}$. Our new astrometric measurements are shown in \autoref{fig:11}, right panel), alongside all previous relative astrometry reported in the literature \citep{Lowrance2000, Guenther2001, Geissler2008, Neuhauser2011, Rameau2013, nogueira2024astrometric} and our calculated relative astrometry from Gaia. 

We observe no significant change in separation or proper motion. The new MIRI MRS measurements for both separation and position angle are consistent with $\eta$ Tel B being a common proper motion companion to $\eta$ Tel A located at or near the apocentre of a long-period orbit. This confirms previous analysis \citep{Neuhauser2011}.

\subsection{Orbit Fitting}
\label{ssec:Borbfit}

To better characterise the orbital properties of $\eta$ Tel B, and to understand any potential companion-disk interactions, it is necessary to first understand the orbit of $\eta$ Tel B. 
We perform an orbital fit for the brown dwarf companion with the Python package \texttt{orbitize!} \citep{Blunt2020}, using relative astrometry and stellar absolute astrometry from the DR3 Hipparcos-Gaia Catalogue of Accelerations (HGCA; \cite{Brandt2021}). Additionally, although HARPS radial velocity (RV) data exists for $\eta$ Tel A \citep{trifonov2020}, the primary shows a high rms RV scatter of $12.805\pm0.007$ km s$^{-1}$ due to its A0V spectral type, youth, and fast rotation. This makes it difficult to obtain meaningful constraints from the RV data. As such, we omit RV data from our orbital fit for the companion.

We run a parallel-tempered Markov-Chain Monte Carlo (MCMC; \cite{ForemanMackey2013, Vousden2016} algorithm with 10 temperatures, 500 walkers and $10^6$ steps, burning the first 100 steps and thinning every 1000 steps; we select these MCMC parameters to maintain consistency with those used by \cite{nogueira2024astrometric}. We set normal priors on the stellar mass ($2.09\pm0.03 M_{\odot}$, \cite{Desidera2021}, and parallax of the system ($\pi=20.6028\pm0.0988$ mas; \cite{GAIA2023}), as well as a uniform prior on the companion mass (0.019--0.048 $M_\odot$, i.e. 20--50 $M_J$, \cite{Neuhauser2011}). Uninformative priors are adopted for all other orbital elements; we use the \texttt{orbitize!} defaults given in \autoref{tab:2}. We calculate posteriors for 9 parameters: semi-major axis $a_B$, eccentricity $e_B$, inclination $i_B$, argument of pericentre $\omega_B$, longitude of ascending node $\Omega_B$, and epoch of pericentre $\tau_B$. \autoref{tab:2} gives the full list of our derived orbital parameters to $1$-$\sigma$ uncertainties. We also do not specify initial positions for the MCMC chains, instead using the \texttt{orbitize!} default, which randomly determines the initial position of the walkers such that they are uniformly distributed across the prior phase space. To test for convergence, we check trace plots and posterior histograms for each parameter.

We obtain best-fit median values of semi-major axis $a_B=142^{+18}_{-11}$ au, eccentricity $e_B=0.50^{+0.1}_{-0.1}$, and inclination $i_B=79^{+5}_{-6}$ degrees. This gives an apocentre distance of $r_{\mathrm{max},B}=213$ au, a pericentre distance of $r_{\mathrm{min},B}=71$ au, and an orbital period $t_B$$\sim$1100 years. A lack of significant change in orbital motion across twenty-five years of observations is therefore reasonable, as we have only observed $\sim$2\% of the companion's total orbit. \autoref{fig:11} shows a sample of 100 potential orbits, as well as the corresponding projected change in separation and position angle for each of these 100 orbits. Due to the long-period nature of $\eta$ Tel B's orbit, it will be difficult to observe any significant changes within the next decade; placing more robust constraints on the companion's orbital parameters may not be possible until several decades from now. 

We note that, while our values for $a$, $e$, and $i$ are in agreement with the orbital parameters inferred by \cite{Neuhauser2011}, our values for $a$ and $e$ differ considerably from those derived by \cite{nogueira2024astrometric} using \texttt{orvara} \citep{Brandt2021}. To investigate the potential reasons for this discrepancy, we perform several additional \texttt{orbitize!} fits using (1) their Gaussian prior on the companion mass of 47$\pm$15 $M_J$, (2) only their relative astrometry data, and (3) their initial distribution values. Corner plots for these additional fits are shown in the \autoref{app:1}. We find that the original derived parameters for $a$ and $e$ remain robust to the change in companion mass prior and the additional Gaia and MIRI MRS relative astrometry points. The fit using the \cite{nogueira2024astrometric} initial distribution values returns a bimodal posterior distribution for $a$ and, to a lesser extent, $e$. The tallers peaks ($a\sim149$ au and $e=0.5$) are consistent with our posteriors but the shorter peaks ($a\sim230$ au and $e\sim0.3$) are consistent with their results. This suggests that the choice of initial position may have some affect on the posteriors. In particular, instead of using a uniform distribution to set the initial position of the walkers for each parameter, \cite{nogueira2024astrometric} use a log-normal distribution for the semi-major axis and a normal distribution for all other orbital values. In difficult cases, such as determining $a$ and $e$ for long-period orbits, this may preferentially concentrate exploration of values to those near the chosen initial values.

However, the fact that \texttt{orbitize!} does not exactly reproduce the posteriors from \cite{nogueira2024astrometric} despite using the same initial distributions, and also consistently has smaller uncertainties than \texttt{orvara}, suggests that some more fundamental difference between the two fitting packages may also be contributing to the different fit outcomes. For example, \texttt{orvara} parametrises $e_B$ as $e_B\sin{\Omega_B}$ and $e_B\cos{\Omega_B}$, whereas \texttt{orbitize!} does not). Further investigation may provide more illuminating information, but as a deep-dive into the workings of both fitting packages is outside the scope of this work, we leave it to future work.

We also note that, since observations of $\eta$ Tel B to date only cover a small fraction of its total orbital period, its astrometric acceleration between Hipparcos and Gaia in the HGCA has a low significance of 1.96-$\sigma$ for two degrees of freedom \citep{Brandt2021}. As such, while fitting for the stellar absolute astrometry is able to constrain the direction of orbital motion, it is unlikely to provide strong constraints on the dynamical mass of the companion. This is reflected in our median companion mass posterior of $M=42\pm14 M_J$, which appears to be largely prior-driven.

In our following analysis, we assume the best-fit orbital parameters described in Table 2; however, we acknowledge that these parameters may be in part due to the fitting package used, and as such also reproduce the analysis using the best-fit parameters from \cite{nogueira2024astrometric}.

\begin{table}
    \centering
    \caption{Median \texttt{orbitize!} Posteriors for $\eta$ Tel B}
    \begin{tabular}{ccc}
        \hline
        Parameter&  Value&  Prior\\
        \hline
        $a_B$ [au]&               $142^{+18}_{-11}$&    log uniform\\
        $e_B$&                    $0.5\pm0.1$& uniform\\
        $i_B$ [$^{\circ}$]&       $79^{+5}_{-6}$&       sine\\
        $\omega_B$ [$^{\circ}$]&  $169^{+23}_{-21}$&    uniform\\
        $\Omega_B$ [$^{\circ}$]&  $169^{+3}_{-2}$&      uniform\\
        $\tau_B$ &                $0.5\pm0.1$&   uniform\\
        \hline
        $r_{min,B}$ [au]&         $71^{+9}_{-6}$&     -\\
        $r_{max,B}$ [au]&         $213^{+27}_{-17}$&    -\\
        $t_B$ [yr]&               $1100^{+230}_{-132}$&  -\\
        \hline
        $M_B$ [$M_J$]&               $35^{+7}_{-8}$&    uniform\\
        $M_A$ [$M_\odot$]& $2.09\pm0.03$&               Gaussian\\
        $\pi$ [mas]& $20.61\pm0.07$&                    Gaussian\\
        \hline
    \end{tabular}
    \label{tab:2}
\end{table}

\section{Discussion}
\label{sec:discussion}

An isolated, nearly edge-on debris disk comprised of planetesimals on circular orbits will feature a symmetric, double-lobed structure. However, dynamical interactions due to the presence of a massive second body in the system can sculpt the structure of the disk, leading to asymmetries \citep{Wyatt1999}.

In the case of the $\eta$ Tel AB system, we find that the eccentricity of the companion's orbit ($e_B=0.50$) corresponds to a pericentre distance of $\sim$71 au from the primary star (see \autoref{tab:1}). 
Given the outer disk's radial extent between $r_{in}\sim$22 and $r_{out}\sim$26 au \citep{Smith2009}, we expect that the companion passes close enough at its pericentre to gravitationally perturb the material within disk over secular timescales.

For low eccentricity orbits, we can expect secular precession to act on a timescale given by (\cite{Wyatt2005}, see Eqs. 7 and 8 therein),

\begin{equation}
    t_{sec}=\frac{6.15\alpha^{-2.5}\bar\alpha^2}{b^{(1)}_{3/2}(\alpha_{B})}\frac{0.651t_{B}}{\mu}
\end{equation}

\noindent For higher eccentricities, the above expression should still give a reasonable estimate. In the case of $\eta$ Tel A and B, the ratio of perturber to disk semi-major axes is $\alpha=a_d/a_B\approx24/142\approx0.17$, with $\bar\alpha=\alpha$ since $a_B>a_{d}$. The Laplace coefficient is $b^{(1)}_{3/2}(\alpha_{B})\approx3\alpha\approx0.56$, and the perturber's orbital period in years is $t_{B}\approx1100$ yrs (see \autoref{tab:2}). We set the ratio of perturber to star masses as $\mu\equiv M_B/M_*=35M_{J}/2.09 M_\odot\approx0.02$, using the companion mass derived in \hyperref[ssec:Borbfit]{Section 5.2}. 

This gives $t_{sec}\approx1$ Myr. Performing the same calculation using the companion orbital parameters derived by \cite{nogueira2024astrometric} provides a similar result of $t_{sec}\approx2$ Myr. Placed into context with the $\beta$ Pic moving group age of $\sim$23 Myr, we should expect that the observed properties of the $\eta$ Tel A disk are consistent with the stable end-product of secular interactions with $\eta$ Tel B, regardless of which fit parameters are used.

We next discuss the predicted disk properties due to dynamical interaction with $\eta$ Tel B, and compare these predictions to the observed MRS data. For our analysis, we consider both our best-fit orbital parameters as well as those derived by \cite{nogueira2024astrometric}.

\subsection{Radial Extent of the Disk}
\label{ssec:radext}

In the case of material orbiting a primary star, with a secondary binary companion acting as a perturber on the material, there should be a critical semi-major axis at which the orbit of the material is stable against gravitational perturbations from the companion. \cite{Holman1999} empirically derive an expression for this critical semi-major axis, $a_c$, as a function of the primary-secondary mass ratio, $\mu$, and the semi-major axis, $a_B$, and eccentricity, $e_B$, of the secondary perturber's orbit:

\begin{multline}
    a_c = [(0.464\pm0.006)\\ + (-0.38\pm0.01)\mu + (-0.631\pm0.034)e_B\\ + (0.586\pm0.061)\mu e_B + (0.15\pm0.041)e_B^2\\ + (-0.198\pm0.074)\mu e_B^2]a_B
\end{multline}

For $\eta$ Tel A and B where $\mu\sim0.02$, we obtain a critial semi-major axis of $a_c=26.2$ au. This is comparable to the inferred radial extent of the disk ($r_{out}=26$ au, \cite{Smith2009}), suggesting that the observed structure of the disk is consistent with truncation due to the orbit of the brown dwarf companion.

For the \cite{nogueira2024astrometric} values of $a_B=218$ au and $e_B=0.34$, we obtain $a_c=57$ au. This is greater than the outer radial extent of the disk reported in the literature. However, we note that, if the apparent increase in radial extent of the disk seen in the MIRI MRS data is real and not an artefact of PSF subtraction (see \autoref{ssec:Adiskspat}, \autoref{fig:6} and \hyperref[fig:7]{7}), then at greatest extent the disk does not seem to exceed past $\sim$60 au. This could be consistent with truncation by an $\eta$ Tel B with a larger semi-major axis of $\sim$220 au.

\subsection{Symmetry of the Disk}
\label{ssec:symm}

Although an axisymmetric, double-lobed structure is expected for an isolated debris disk, secular perturbations due to the gravitational influence of a second, eccentric body in the system can force the orbit of dust within the disk to become likewise eccentric. This shifts the symmetry of the disk away from the star, resulting in an observable `pericentre glow' as dust at the forced pericentre of the disk is heated by increased proximity to the stellar host \citep{Wyatt1999}.

Since a particle's forced eccentricity $e_f$ depends only on the eccentricity of the perturber's orbit along with the ratio of its semimajor axis to that of the perturber (\cite{Wyatt1999}, Eq. 39), we estimate the forced eccentricity due to $\eta$ Tel B as follows:

\begin{equation}
    e_f \simeq \frac{5}{4} \frac{a_{d}}{a_{B}} \times e_B = \frac{5}{4} \frac{24~\mathrm{au}}{143~\mathrm{au}} \times 0.50  \approx 0.1
\end{equation}

Here, the perturber semi-major axis, $a_B$, and orbital eccentricity, $e_B$, come from our derived orbital parameters for $\eta$ Tel B. We again take the mean planetesimal belt distance from the star, $a_{d}$, to be 24 au \citep{Smith2009}.
This gives us planetesimal belt apocentre and pericentre distances of 26.5 au and 21.5 au respectively. 
We then calculate the grain temperature of the dust at both these distances using the following \citep{Chen2001}:

\begin{equation}
    T_{gr} = 0.707T_*\sqrt{\frac{R_*}{D_{gr}}}
    \label{eq:2}
\end{equation}

\begin{figure*}[ht]
    \centering
    \includegraphics[width=\textwidth]{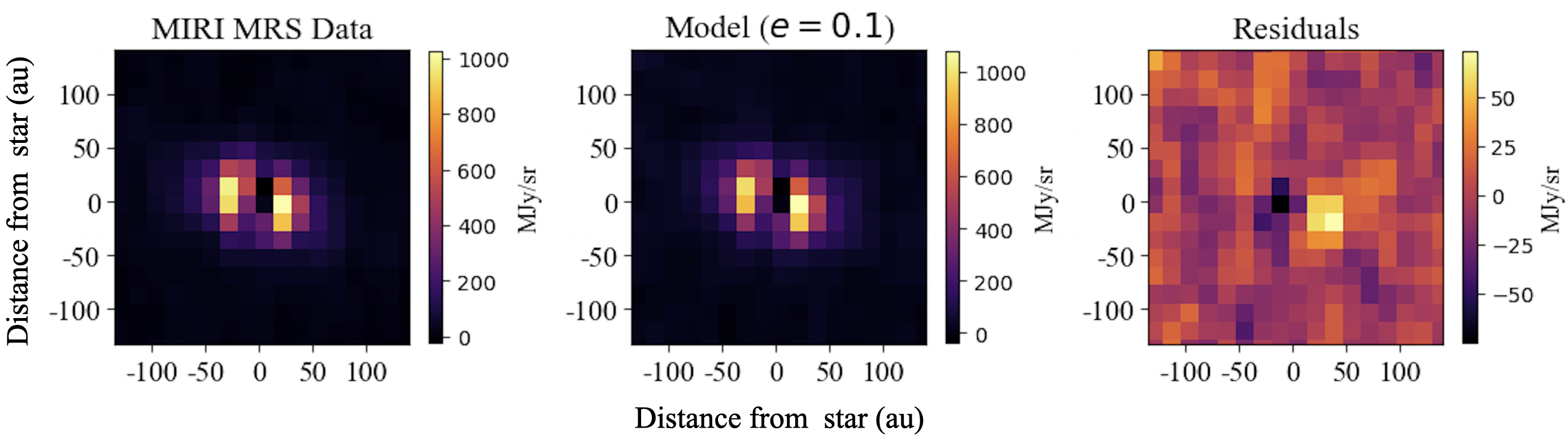}
    \caption{(\textit{Left}) 18 $\mu m$ slice of MIRI MRS data, following peak-scaled PSF subtraction. Both lobes appear similar in brightness (\textit{Centre}) Model of disk with $e_f=0.1$ that has been convolved with the MIRI MRS PSF at 18 $\mu m$, then PSF-subtracted. The pericentre lobe appears slightly brighter. (\textit{Right}) Residuals after subtracting the MIRI MRS data slice from the model image, showing remnant structure in the pericentre lobe.}
    \label{fig:12}
\end{figure*}

\noindent where $D_{gr}$ is the distance of the grains from the star, $T_*=9700$ K, and $R_*=1.7R_\odot$ \citep{Chen14}. This gives a grain temperature of 118 K at the disk's apocentre, and 132 K at the disk's pericentre. Taking the ratio of blackbody flux densities across $\lambda_c$ for each MRS sub-band, we estimate an expected brightness asymmetry of $96\%$ at 18 $\mu m$, which we then divide by a factor of $1+e_f$ to account for particle bunching at apocentre \citep{Pan2016}. This gives us a final expected brightness asymmetry of $77\%$ at 18 $\mu m$, with the pericentre lobe being brighter. It is worth noting that this may be an overestimation of the brightness asymmetry due to the assumption that the dust is a blackbody; in reality, the temperature of the dust may be hotter. However, given the cool dust component temperature of 127 K from dust modelling (see \hyperref[ssec:Adustmod]{Section 3.2}), this would not be a large correction. Likewise, a higher $T_\mathrm{eff}$ for the primary would give rise to hotter grain temperatures and a smaller brightness asymmetry.

Repeating the above calculations using the best-fit orbital parameters from \cite{nogueira2024astrometric} produces a forced eccentricity of $e_f=0.04$, which should produce a 18 $\mu m$ pericentre brightness asymmetry of $\sim$30$\%$.

Both scalings of the PSF-subtracted MRS data cubes, however, appear largely axisymmetric (\autoref{fig:6} and \hyperref[fig:7]{7}), which is inconsistent with expectations of an observable brightness asymmetry. \autoref{fig:12} compares an 18 $\mu m$ slice of the MIRI MRS data (after peak-scaled PSF subtraction) to a model of the disk with $e_f=0.1$. Subtracting the MIRI MRS data from the model image reveals residuals that indicate the model is brighter in the pericentre lobe than the data; i.e. the data is less asymmetric than expected from consideration of the system's dynamics. To perform a more rigorous check for potentially fainter asymmetries, we apply angular differential imaging (ADI) to each collapsed sub-band image of the disk. This is done by rotating the image 180$^{\circ}$ and subtracting it from the unrotated image. We also perform the same ADI on the N Car data in order to check whether any potential structures are due to instrument effects. Although we find some asymmetric structure, it is inconsistent across wavelengths and, more critically, appears in both sets of observations. This indicates that these structures are likely caused by the instrument rather than any real physical asymmetry in the $\eta$ Tel disk. Thus, we find that the $\eta$ Tel A disk is essentially axisymmetric, contrary to our expectation of an observable disk asymmetry due to gravitational perturbation by $\eta$ Tel B.

\subsection{Mutual Inclination of the Disk and Companion}
\label{ssec:mutinc}

Secular precession induced by the orbit of $\eta$ Tel B should cause the orbital planes of the disk and the companion to become aligned; i.e., for a 23 Myr system, we should expect to observe an aligned mutual inclination between the disk and the companion, even if the two were initially misaligned.

The mutual inclination of the disk and $\eta$ Tel B can be calculated using,

\begin{equation}
    \cos{i_m}=\cos{i_d}\cos{i_B}+\sin{i_d}\sin{i_B}\cos{(\Omega_B-\Omega_d)}
\end{equation}

\noindent where the $i$ terms are inclinations relative to the sky plane and the $\Omega$ terms are the longitudes of ascending node. For disk and companion parameters of $i_d=90\pm20$ deg and $\Omega_d=172\pm1$ deg \citep{Smith2009}, and $i_B=79^{+5}_{-6}$ deg and $\Omega_B=169^{+3}_{-2}$ deg (see \autoref{tab:2}, \hyperref[ssec:Borbfit]{Section 5.2}), we obtain a mutual inclination of $i_m\sim$$11^{+15}_{-14}$ deg. Thus, we find that the disk and the companion may potentially be misaligned, contrary to expectation. However, further modelling, particularly of the disk's parameters, is needed to improve uncertainties before we can determine whether or not the disk is truly misaligned with the companion's orbit.

\subsection{An Additional Interior Planet?}
\label{ssec:pldyn}

The absence of compelling evidence for asymmetry in the debris disk, and its potential misalignment with the companion, presents an intriguing puzzle that deserves explanation. Motivated by the work of \cite{Farhat2023} concerning the HD 106906 debris disk, which is perturbed by both an exterior companion and an inner stellar binary, we propose that our observations could be explained by the presence of additional perturbing masses in the $\eta$ Tel system. In principle, these masses may include either single or multiple planets interior to the disk, and/or the self-gravitational effects of the disk itself, if massive enough. The reasoning behind this is that the presence of such additional masses could counteract the gravitational effects of $\eta$ Tel B on the debris disk, explaining the identified discrepancies (\hyperref[ssec:symm]{Section 6.2}, \hyperref[ssec:mutinc]{6.3}).

We consider what may be the simplest scenario: an additional, single, yet-undetected planet on a circular orbit completely interior to, and coplanar with, the debris disk (assumed to be massless). It is then possible to constrain the mass $m_{pl}$ and semimajor axis $a_{pl}$ of such a planet using the so-called ``Laplace radius'' \citep{Tremaine2009, Farhat2023}. The Laplace radius, denoted by $r_L$, describes the location where the gravitational perturbations experienced by a planetesimal due to both the inner and outer companions are equal and cancel out. Thus, for a given system, planetesimal dynamics interior (exterior) to $r_L$ will be dominated by the inner (outer) companion, with planetesimals lying in the dominant companion’s orbital plane. The Laplace radius, in the limit of $m_{pl}<<M_*$, can be written as follows (see equation (1) in \cite{Farhat2023}):

\begin{equation}
    r_L^5 = a_B^3 a_{pl}^2 \frac{m_{pl}}{M_B}(1-e_B^2)^{\frac{3}{2}}
    \label{eq:lagrange}
\end{equation}

\begin{figure}[ht]
    \centering
    \includegraphics[width=0.45\textwidth]{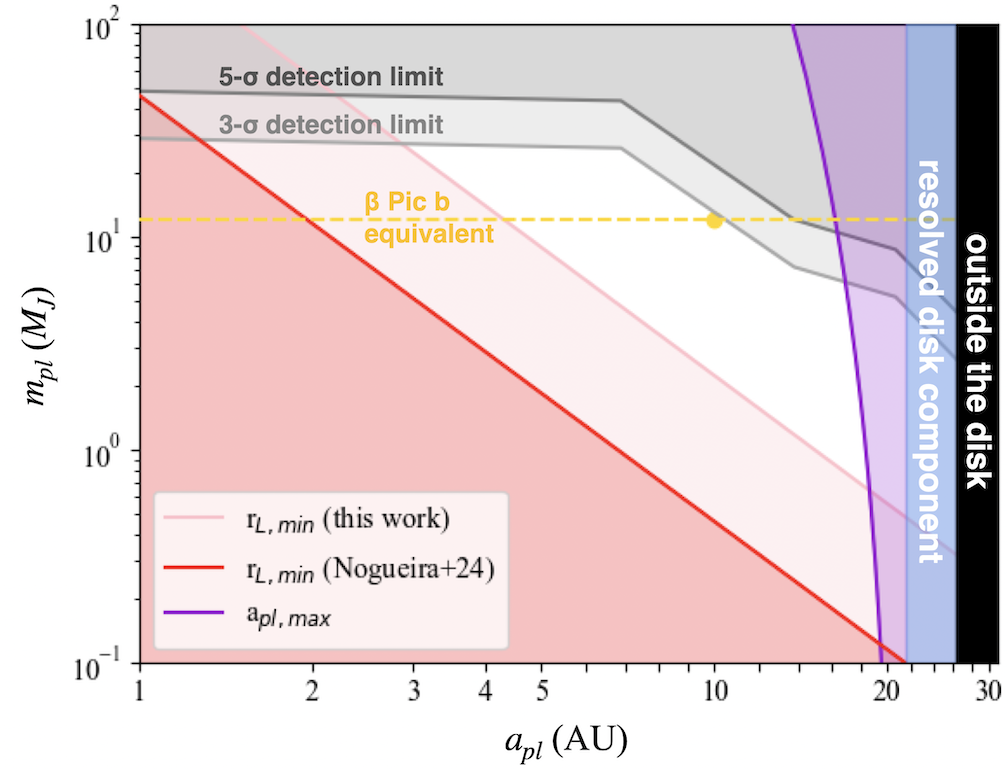}
    \caption{Mass--semi-major axis constraints for a potentially undetected perturbing planet within the disk. The blank white region in the centre indicates the possible parameter space for an undetected, inner planetary perturber on a circular orbit within the disk plane. The pink lines correspond to the minimum Laplace radius ($r_{out}$ of the disk, see \autoref{eq:lagrange}) and sets the minimum possible mass for the planet at a given semimajor axis $a_{pl}$. The purple line shows the upper bound on $a_{p}$ set by the MMR overlap argument (\autoref{eq:mmr}). The grey lines exclude planets within MIRI MRS 3- and 5-$\sigma$ detection limits. Note also that the region outside the disk, marked in black, is unstable due to perturbations from $\eta$ Tel B (\hyperref[ssec:radext]{Section 6.1}).}
    \label{fig:13}
\end{figure}

Since the observed disk structure seems to be inconsistent with that expected based on secular perturbations due to $\eta$ Tel B alone (\hyperref[ssec:symm]{Sections 6.2} and \hyperref[ssec:mutinc]{6.3}), in \autoref{eq:lagrange} we set the minimum Laplace radius to be the disk’s outermost radius (i.e., $r_L\geq r_{out}\approx$ 26 au, \cite{Smith2009}), and solve for $m_{pl}$ as a function of $a_{pl}$. The results are shown in \autoref{fig:13}; a planet whose parameters lie above the light pink line (i.e. values of $m_{pl}$ and $a_{pl}$ for which $r_L \geq r_{out}$) could maintain the disk’s axisymmetry and its misalignment with the outer companion. We note that using the \cite{nogueira2024astrometric} orbital parameters to solve for $m_{pl}$ as a function of $a_{pl}$ (shown in dark pink in \autoref{fig:13}, where again we have set $r_L=r_{out}=26$ au) results in a larger possible parameter space for an undetected interior planet.

We further constrain the possible parameter space as follows. First, we use MIRI MRS 3- and 5-$\sigma$ contrast curves to determine instrument detection limits for potential companions. This allows us to rule out the region of the parameter space as shown using the grey lines in \autoref{fig:13}.
For comparison, we also calculate the expected contrast for a $\beta$ Pic b-like planet around $\eta$ Tel using the following equation:
\begin{equation}
    f_1 = (\frac{d_2}{d_1})^2 \times f_2
\end{equation}
\noindent We use the $\beta$ Pic b atmosphere model from \cite{Worthen2024} to obtain its flux $f_2$ at $\lambda_c=5.3$, 6.2 and 7.1 $\mu m$ (the central wavelengths of sub-bands 1A--1C respectively). The distances of $\eta$ Tel and $\beta$ Pic are $d_1=47.7$ and $d_2=19.44$ pc respectively. We then divide $f_2$ by the stellar flux at each wavelength to obtain the final contrasts, finding that MIRI MRS should be able to detect a $\beta$ Pic b-like planet of $m_{pl}\sim12 M_J$ at $\geq$18 au (within 5-$\sigma$, and at $\geq$ 12 au within 3-$\sigma$).

Second, assuming that the inner edge of the disk is carved by the overlap of first-order mean motion resonances due to the planet (e.g. \cite{Pearce2024} and references therein), the planet’s semimajor axis cannot exceed $a_{p}=a_{d} - \Delta a_{p}$. Here, $\Delta a_{p}$ is the half-width of the chaotic zone around the planetary orbit given by the following expression \citep{Wisdom1980}:

\begin{equation}
    \Delta a_p \approx 1.3(\frac{m_p}{M_*+m_p})^{2/7} a_p
    \label{eq:mmr}
\end{equation}

This is shown in \autoref{fig:13} using a purple curve.

Given these bounds, we are able to rule out certain areas of the planet’s possible mass and semimajor axis, as summarised in \autoref{fig:13}. The central white region therein represents the allowed parameter space for an undetected planet interior to the disk. Looking at \autoref{fig:13} it is evident that a planet of mass $\sim$0.7--30 $M_J$ and semimajor axis $\sim$3--19 au may be responsible for the observed disk structure (alternatively, 0.15--40 $M_J$ between 1.5--20 au if using companion orbital parameters from \cite{nogueira2024astrometric}). That being said, however, we stress that our aim here is not to offer a quantitative prediction but rather to highlight that the observed disk structure is a plausible consequence of the presence of an additional planet.

This is because several additional factors may influence our predictions, which we discuss below. First, the Laplace radius of Equation (7) does not account for potential interactions between the inner and outer perturbers, and is instead derived assuming $a_{pl} << a_B$. Second, in our calculations, we do not account for non-gravitational forces. This is fairly reasonable for mm-sized or larger grains; however, the MIRI MRS data traces $\mu m$-sized grains which are subject to non-gravitational forces such as radiation pressure and gas drag. Given the system’s relatively young age, it is possible for the disk to contain significant amount of gas (see, however, \cite{Iglesias2023}) which can affect the dust dynamics, such as through migration and the damping of orbital eccentricities/inclinations \citep{Takeuchi2001}. If this is the case, then a less massive planet than that identified in \autoref{fig:13} would instead be required to produce the same observed structure. Third, we assume that the debris disk is massless, neglecting its (self-)gravitational effects. However, if the disk is massive enough, it may suppress planetesimal eccentricities forced by the eccentric companion \citep{Sefilian2024}, affecting our inferences. In the extreme case, the disk self-gravity alone, without an additional interior planet, can potentially explain the observed disk structure \citep{Batygin2011, Sefilian2024}. Regardless, the disk self-gravity, even if not dominant, may well affect our planetary inferences by forcing an inward shift in the Laplace radius (see \cite{Farhat2023, Sefilian2021}). Finally, it is important to acknowledge that the inferred parameter space is contingent upon accurate knowledge of the outer companion’s orbital parameters. Any updates or improvements to these orbital parameters may necessitate revisions to \autoref{fig:13}.

\section{Conclusions}
\label{sec:concs}

As part of GTO Program 1294, we present MIRI MRS observations of the $\eta$ Telescopii system. Our main findings are:

\begin{itemize}
    \item We detect an infrared excess in the spectrum of $\eta$ Tel A, indicating the presence of thermal emission from circumstellar dust. We recover the 10 $\mu m$ silicate feature and discover a new broad 20 $\mu m$ silicate feature. Dust modelling suggests the continuum is best-fit by two different grain populations at 319 K and 127 K, with the 10 and 20 $\mu m$ silicate features arising due to the presence of large amorphous grains.
    \item We detect the brown-dwarf companion $\eta$ Tel B at a separation of 4" in MRS sub-bands 3A to 4A. We calculate a new epoch of astrometry for $\eta$ Tel B, with $\rho =$4199$\pm$15 mas and PA $=$ 167.36$\pm$0.19$^{\circ}$. Our measurements extend the baseline of astrometric measurements to 25 years. We detect no significant change in orbital motion.
    \item We derive the orbit of $\eta$ Tel B using relative astrometry and obtain the orbital parameters $a_B=142^{+18}_{-11}$ au, $e_B=0.50\pm0.1$, and $i_B=79^{+5}_{-6}$ degrees. This gives an orbital period of $t_B\sim1100$ years. We find that, for our apocentre distance of 214 au, the companion's current location at 209 au validates previous literature suggesting the companion is located at or near apocentre of a long-period orbit.
    \item We present the first 11--21 $\mu m$ spectrum of $\eta$ Tel B. We do not detect an infrared-excess for the object. We perform atmospheric grid model fitting to obtain the following parameters for $\eta$ Tel B: $T_\mathrm{eff,B}=2830^{+20}_{-30}$ K, $log(g)=4.3^{+0.1}_{-0.2}$, $R=2.28\pm0.03 R_J$, $log L_B/L_{\odot}=-2.48\pm0.01$, $M_B=42\pm14 M_{J}$.
    \item Using PSF subtraction, we spatially resolve the debris disk around $\eta$ Tel A from 9.4--26.05 $\mu m$. We find that the disk has an axisymmetric double-lobed structure across the MRS wavelength range. This is inconsistent with the expected 77\% brightness asymmetry at 18 $\mu m$ due to secular perturbations from $\eta$ Tel B, assuming our median orbital parameters for the companion.
    \item The disk's axisymmetric structure 
    and potential misalignment with the companion may be due to the presence of another mass in the system that is large enough to dominate over secular precessional effects induced by $\eta$ Tel B. For the case of a single, yet-undetected planet, we constrain its mass to be between $\sim$0.7--30 $M_J$ with a semi-major axis within the $\sim$3--19 au range (\autoref{fig:13}).
\end{itemize}

We thank the anonymous referee for their helpful and insightful comments. YC and CC acknowledge that this work is based [in part] on observations made with the NASA/ESA/CSA James Webb Space Telescope. The data were obtained from the Mikulski Archive for Space Telescopes at the Space Telescope Science Institute, which is operated by the Association of Universities for Research in Astronomy, Inc., under NASA contract NAS 5-03127 for JWST. These observations are associated with program \#1294. Support for program \#1294 was provided by NASA through a grant from the Space Telescope Science Institute, which is operated by the Association of Universities for Research in Astronomy, Inc., under NASA contract NAS 5-03127. A.A.S. is supported by the Alexander von Humboldt Foundation through a Humboldt Research Fellowship for postdoctoral researchers.

\software{
This research has made use of the following software projects:
    \href{https://astropy.org/}{Astropy} \citep{astropy2013, astropy18, astropy2022},
    \href{https://matplotlib.org/}{Matplotlib} \citep{matplotlib07},
    \href{http://www.numpy.org/}{NumPy} and \href{https://scipy.org/}{SciPy} \citep{numpy07},
    \href{https://jwst-pipeline.readthedocs.io/en/latest/}{JWST DataPipeline} (Bushouse et al. 2023),
    \href{https://github.com/farhanferoz/MultiNest}{MultiNest} \citep{Feroz09},
    \href{https://orbitize.readthedocs.io/en/latest/}{orbitize!} \citep{Blunt2020},
    \href{http://johannesbuchner.github.io/PyMultiNest/pymultinest.html}{PyMultiNest} \citep{buchner14},
    \href{https://species.readthedocs.io/en/latest/}{species} \citep{Stolker2020},
    and
    the NASA Astrophysics Data System.
}


\bibliographystyle{aasjournal}
\bibliography{bib}

\begin{thebibliography}{}
\expandafter\ifx\csname natexlab\endcsname\relax\def\natexlab#1{#1}\fi
\providecommand{\url}[1]{\href{#1}{#1}}
\providecommand{\dodoi}[1]{doi:~\href{http://doi.org/#1}{\nolinkurl{#1}}}
\providecommand{\doeprint}[1]{\href{http://ascl.net/#1}{\nolinkurl{http://ascl.net/#1}}}
\providecommand{\doarXiv}[1]{\href{https://arxiv.org/abs/#1}{\nolinkurl{https://arxiv.org/abs/#1}}}

\bibitem[{{Allard} {et~al.}(2011){Allard}, {Homeier}, \& {Freytag}}]{allard2011}
{Allard}, F., {Homeier}, D., \& {Freytag}, B. 2011, in Astronomical Society of the Pacific Conference Series, Vol. 448, 16th Cambridge Workshop on Cool Stars, Stellar Systems, and the Sun, ed. C.~{Johns-Krull}, M.~K. {Browning}, \& A.~A. {West}, 91, \dodoi{10.48550/arXiv.1011.5405}

\bibitem[{{Apai}(2013)}]{Apai2013}
{Apai}, D. 2013, Astronomische Nachrichten, 334, 57, \dodoi{10.1002/asna.201211780}

\bibitem[{{Argyriou} {et~al.}(2023){Argyriou}, {Glasse}, {Law}, {Labiano}, {{\'A}lvarez-M{\'a}rquez}, {Patapis}, {Kavanagh}, {Gasman}, {Mueller}, {Larson}, {Vandenbussche}, {Glauser}, {Royer}, {Dicken}, {Harkett}, {Sargent}, {Engesser}, {Jones}, {Kendrew}, {Noriega-Crespo}, {Brandl}, {Rieke}, {Wright}, {Lee}, \& {Wells}}]{argyriou2023}
{Argyriou}, I., {Glasse}, A., {Law}, D.~R., {et~al.} 2023, arXiv e-prints, arXiv:2303.13469, \dodoi{10.48550/arXiv.2303.13469}

\bibitem[{{Astropy Collaboration} {et~al.}(2013){Astropy Collaboration}, {Robitaille}, {Tollerud}, {Greenfield}, {Droettboom}, {Bray}, {Aldcroft}, {Davis}, {Ginsburg}, {Price-Whelan}, {Kerzendorf}, {Conley}, {Crighton}, {Barbary}, {Muna}, {Ferguson}, {Grollier}, {Parikh}, {Nair}, {Unther}, {Deil}, {Woillez}, {Conseil}, {Kramer}, {Turner}, {Singer}, {Fox}, {Weaver}, {Zabalza}, {Edwards}, {Azalee Bostroem}, {Burke}, {Casey}, {Crawford}, {Dencheva}, {Ely}, {Jenness}, {Labrie}, {Lim}, {Pierfederici}, {Pontzen}, {Ptak}, {Refsdal}, {Servillat}, \& {Streicher}}]{astropy2013}
{Astropy Collaboration}, {Robitaille}, T.~P., {Tollerud}, E.~J., {et~al.} 2013, \aap, 558, A33, \dodoi{10.1051/0004-6361/201322068}

\bibitem[{{Astropy Collaboration} {et~al.}(2022){Astropy Collaboration}, {Price-Whelan}, {Lim}, {Earl}, {Starkman}, {Bradley}, {Shupe}, {Patil}, {Corrales}, {Brasseur}, {N{\"o}the}, {Donath}, {Tollerud}, {Morris}, {Ginsburg}, {Vaher}, {Weaver}, {Tocknell}, {Jamieson}, {van Kerkwijk}, {Robitaille}, {Merry}, {Bachetti}, {G{\"u}nther}, {Aldcroft}, {Alvarado-Montes}, {Archibald}, {B{\'o}di}, {Bapat}, {Barentsen}, {Baz{\'a}n}, {Biswas}, {Boquien}, {Burke}, {Cara}, {Cara}, {Conroy}, {Conseil}, {Craig}, {Cross}, {Cruz}, {D'Eugenio}, {Dencheva}, {Devillepoix}, {Dietrich}, {Eigenbrot}, {Erben}, {Ferreira}, {Foreman-Mackey}, {Fox}, {Freij}, {Garg}, {Geda}, {Glattly}, {Gondhalekar}, {Gordon}, {Grant}, {Greenfield}, {Groener}, {Guest}, {Gurovich}, {Handberg}, {Hart}, {Hatfield-Dodds}, {Homeier}, {Hosseinzadeh}, {Jenness}, {Jones}, {Joseph}, {Kalmbach}, {Karamehmetoglu}, {Ka{\l}uszy{\'n}ski}, {Kelley}, {Kern}, {Kerzendorf}, {Koch}, {Kulumani}, {Lee}, {Ly}, {Ma}, {MacBride}, {Maljaars}, {Muna}, {Murphy}, {Norman},
  {O'Steen}, {Oman}, {Pacifici}, {Pascual}, {Pascual-Granado}, {Patil}, {Perren}, {Pickering}, {Rastogi}, {Roulston}, {Ryan}, {Rykoff}, {Sabater}, {Sakurikar}, {Salgado}, {Sanghi}, {Saunders}, {Savchenko}, {Schwardt}, {Seifert-Eckert}, {Shih}, {Jain}, {Shukla}, {Sick}, {Simpson}, {Singanamalla}, {Singer}, {Singhal}, {Sinha}, {Sip{\H{o}}cz}, {Spitler}, {Stansby}, {Streicher}, {{\v{S}}umak}, {Swinbank}, {Taranu}, {Tewary}, {Tremblay}, {de Val-Borro}, {Van Kooten}, {Vasovi{\'c}}, {Verma}, {de Miranda Cardoso}, {Williams}, {Wilson}, {Winkel}, {Wood-Vasey}, {Xue}, {Yoachim}, {Zhang}, {Zonca}, \& {Astropy Project Contributors}}]{astropy2022}
{Astropy Collaboration}, {Price-Whelan}, A.~M., {Lim}, P.~L., {et~al.} 2022, \apj, 935, 167, \dodoi{10.3847/1538-4357/ac7c74}

\bibitem[{{Backman} \& {Paresce}(1993)}]{Backman1993}
{Backman}, D.~E., \& {Paresce}, F. 1993, in Protostars and Planets III, ed. E.~H. {Levy} \& J.~I. {Lunine}, 1253

\bibitem[{{Bailey} {et~al.}(2014){Bailey}, {Meshkat}, {Reiter}, {Morzinski}, {Males}, {Su}, {Hinz}, {Kenworthy}, {Stark}, {Mamajek}, {Briguglio}, {Close}, {Follette}, {Puglisi}, {Rodigas}, {Weinberger}, \& {Xompero}}]{Bailey2014}
{Bailey}, V., {Meshkat}, T., {Reiter}, M., {et~al.} 2014, \apjl, 780, L4, \dodoi{10.1088/2041-8205/780/1/L4}

\bibitem[{{Baraffe} {et~al.}(2003){Baraffe}, {Chabrier}, {Allard}, \& {Hauschildt}}]{Baraffe2003}
{Baraffe}, I., {Chabrier}, G., {Allard}, F., \& {Hauschildt}, P. 2003, in Brown Dwarfs, ed. E.~{Mart{\'\i}n}, Vol. 211, 41

\bibitem[{{Batygin} {et~al.}(2011){Batygin}, {Morbidelli}, \& {Tsiganis}}]{Batygin2011}
{Batygin}, K., {Morbidelli}, A., \& {Tsiganis}, K. 2011, \aap, 533, A7, \dodoi{10.1051/0004-6361/201117193}

\bibitem[{{Blunt} {et~al.}(2017){Blunt}, {Nielsen}, {De Rosa}, {Konopacky}, {Ryan}, {Wang}, {Pueyo}, {Rameau}, {Marois}, {Marchis}, {Macintosh}, {Graham}, {Duch{\^e}ne}, \& {Schneider}}]{Blunt2017}
{Blunt}, S., {Nielsen}, E.~L., {De Rosa}, R.~J., {et~al.} 2017, \aj, 153, 229, \dodoi{10.3847/1538-3881/aa6930}

\bibitem[{{Blunt} {et~al.}(2020){Blunt}, {Wang}, {Angelo}, {Ngo}, {Cody}, {De Rosa}, {Graham}, {Hirsch}, {Nagpal}, {Nielsen}, {Pearce}, {Rice}, \& {Tejada}}]{Blunt2020}
{Blunt}, S., {Wang}, J.~J., {Angelo}, I., {et~al.} 2020, \aj, 159, 89, \dodoi{10.3847/1538-3881/ab6663}

\bibitem[{{Bonnefoy} {et~al.}(2014){Bonnefoy}, {Chauvin}, {Lagrange}, {Rojo}, {Allard}, {Pinte}, {Dumas}, \& {Homeier}}]{Bonnefoy2014}
{Bonnefoy}, M., {Chauvin}, G., {Lagrange}, A.~M., {et~al.} 2014, \aap, 562, A127, \dodoi{10.1051/0004-6361/201118270}

\bibitem[{{Brandt}(2021)}]{Brandt2021}
{Brandt}, T.~D. 2021, \apjs, 254, 42, \dodoi{10.3847/1538-4365/abf93c}

\bibitem[{{Buchner}(2021)}]{Buchner2021}
{Buchner}, J. 2021, The Journal of Open Source Software, 6, 3001, \dodoi{10.21105/joss.03001}

\bibitem[{{Buchner} {et~al.}(2014){Buchner}, {Georgakakis}, {Nandra}, {Hsu}, {Rangel}, {Brightman}, {Merloni}, {Salvato}, {Donley}, \& {Kocevski}}]{buchner14}
{Buchner}, J., {Georgakakis}, A., {Nandra}, K., {et~al.} 2014, \aap, 564, A125, \dodoi{10.1051/0004-6361/201322971}

\bibitem[{{Chen} \& {Jura}(2001)}]{Chen2001}
{Chen}, C.~H., \& {Jura}, M. 2001, \apjl, 560, L171, \dodoi{10.1086/324057}

\bibitem[{{Chen} {et~al.}(2014){Chen}, {Mittal}, {Kuchner}, {Forrest}, {Lisse}, {Manoj}, {Sargent}, \& {Watson}}]{Chen14}
{Chen}, C.~H., {Mittal}, T., {Kuchner}, M., {et~al.} 2014, \apjs, 211, 25, \dodoi{10.1088/0067-0049/211/2/25}

\bibitem[{{Chen} {et~al.}(2006){Chen}, {Sargent}, {Bohac}, {Kim}, {Leibensperger}, {Jura}, {Najita}, {Forrest}, {Watson}, {Sloan}, \& {Keller}}]{Chen2006}
{Chen}, C.~H., {Sargent}, B.~A., {Bohac}, C., {et~al.} 2006, \apjs, 166, 351, \dodoi{10.1086/505751}

\bibitem[{{Cutri} {et~al.}(2003){Cutri}, {Skrutskie}, {van Dyk}, {Beichman}, {Carpenter}, {Chester}, {Cambresy}, {Evans}, {Fowler}, {Gizis}, {Howard}, {Huchra}, {Jarrett}, {Kopan}, {Kirkpatrick}, {Light}, {Marsh}, {McCallon}, {Schneider}, {Stiening}, {Sykes}, {Weinberg}, {Wheaton}, {Wheelock}, \& {Zacarias}}]{cutri2003}
{Cutri}, R.~M., {Skrutskie}, M.~F., {van Dyk}, S., {et~al.} 2003, VizieR Online Data Catalog, II/246

\bibitem[{{Desidera} {et~al.}(2021){Desidera}, {Chauvin}, {Bonavita}, {Messina}, {LeCoroller}, {Schmidt}, {Gratton}, {Lazzoni}, {Meyer}, {Schlieder}, {Cheetham}, {Hagelberg}, {Bonnefoy}, {Feldt}, {Lagrange}, {Langlois}, {Vigan}, {Tan}, {Hambsch}, {Millward}, {Alcal{\'a}}, {Benatti}, {Brandner}, {Carson}, {Covino}, {Delorme}, {D'Orazi}, {Janson}, {Rigliaco}, {Beuzit}, {Biller}, {Boccaletti}, {Dominik}, {Cantalloube}, {Fontanive}, {Galicher}, {Henning}, {Lagadec}, {Ligi}, {Maire}, {Menard}, {Mesa}, {M{\"u}ller}, {Samland}, {Schmid}, {Sissa}, {Turatto}, {Udry}, {Zurlo}, {Asensio-Torres}, {Kopytova}, {Rickman}, {Abe}, {Antichi}, {Baruffolo}, {Baudoz}, {Baudrand}, {Blanchard}, {Bazzon}, {Buey}, {Carbillet}, {Carle}, {Charton}, {Cascone}, {Claudi}, {Costille}, {Deboulb{\'e}}, {De Caprio}, {Dohlen}, {Fantinel}, {Feautrier}, {Fusco}, {Gigan}, {Giro}, {Gisler}, {Gluck}, {Hubin}, {Hugot}, {Jaquet}, {Kasper}, {Madec}, {Magnard}, {Martinez}, {Maurel}, {Le Mignant}, {M{\"o}ller-Nilsson}, {Llored}, {Moulin}, {Orign{\'e}},
  {Pavlov}, {Perret}, {Petit}, {Pragt}, {Puget}, {Rabou}, {Ramos}, {Rigal}, {Rochat}, {Roelfsema}, {Rousset}, {Roux}, {Salasnich}, {Sauvage}, {Sevin}, {Soenke}, {Stadler}, {Suarez}, {Weber}, \& {Wildi}}]{Desidera2021}
{Desidera}, S., {Chauvin}, G., {Bonavita}, M., {et~al.} 2021, \aap, 651, A70, \dodoi{10.1051/0004-6361/202038806}

\bibitem[{{Farhat} {et~al.}(2023){Farhat}, {Sefilian}, \& {Touma}}]{Farhat2023}
{Farhat}, M.~A., {Sefilian}, A.~A., \& {Touma}, J.~R. 2023, \mnras, 521, 2067, \dodoi{10.1093/mnras/stad316}

\bibitem[{{Feroz} {et~al.}(2009){Feroz}, {Hobson}, \& {Bridges}}]{Feroz09}
{Feroz}, F., {Hobson}, M.~P., \& {Bridges}, M. 2009, \mnras, 398, 1601, \dodoi{10.1111/j.1365-2966.2009.14548.x}

\bibitem[{{Foreman-Mackey} {et~al.}(2013){Foreman-Mackey}, {Hogg}, {Lang}, \& {Goodman}}]{ForemanMackey2013}
{Foreman-Mackey}, D., {Hogg}, D.~W., {Lang}, D., \& {Goodman}, J. 2013, \pasp, 125, 306, \dodoi{10.1086/670067}

\bibitem[{{Gaia Collaboration} {et~al.}(2023){Gaia Collaboration}, {Vallenari}, {Brown}, {Prusti}, {de Bruijne}, {Arenou}, {Babusiaux}, {Biermann}, {Creevey}, {Ducourant}, {Evans}, {Eyer}, {Guerra}, {Hutton}, {Jordi}, {Klioner}, {Lammers}, {Lindegren}, {Luri}, {Mignard}, {Panem}, {Pourbaix}, {Randich}, {Sartoretti}, {Soubiran}, {Tanga}, {Walton}, {Bailer-Jones}, {Bastian}, {Drimmel}, {Jansen}, {Katz}, {Lattanzi}, {van Leeuwen}, {Bakker}, {Cacciari}, {Casta{\~n}eda}, {De Angeli}, {Fabricius}, {Fouesneau}, {Fr{\'e}mat}, {Galluccio}, {Guerrier}, {Heiter}, {Masana}, {Messineo}, {Mowlavi}, {Nicolas}, {Nienartowicz}, {Pailler}, {Panuzzo}, {Riclet}, {Roux}, {Seabroke}, {Sordo}, {Th{\'e}venin}, {Gracia-Abril}, {Portell}, {Teyssier}, {Altmann}, {Andrae}, {Audard}, {Bellas-Velidis}, {Benson}, {Berthier}, {Blomme}, {Burgess}, {Busonero}, {Busso}, {C{\'a}novas}, {Carry}, {Cellino}, {Cheek}, {Clementini}, {Damerdji}, {Davidson}, {de Teodoro}, {Nu{\~n}ez Campos}, {Delchambre}, {Dell'Oro}, {Esquej},
  {Fern{\'a}ndez-Hern{\'a}ndez}, {Fraile}, {Garabato}, {Garc{\'\i}a-Lario}, {Gosset}, {Haigron}, {Halbwachs}, {Hambly}, {Harrison}, {Hern{\'a}ndez}, {Hestroffer}, {Hodgkin}, {Holl}, {Jan{\ss}en}, {Jevardat de Fombelle}, {Jordan}, {Krone-Martins}, {Lanzafame}, {L{\"o}ffler}, {Marchal}, {Marrese}, {Moitinho}, {Muinonen}, {Osborne}, {Pancino}, {Pauwels}, {Recio-Blanco}, {Reyl{\'e}}, {Riello}, {Rimoldini}, {Roegiers}, {Rybizki}, {Sarro}, {Siopis}, {Smith}, {Sozzetti}, {Utrilla}, {van Leeuwen}, {Abbas}, {{\'A}brah{\'a}m}, {Abreu Aramburu}, {Aerts}, {Aguado}, {Ajaj}, {Aldea-Montero}, {Altavilla}, {{\'A}lvarez}, {Alves}, {Anders}, {Anderson}, {Anglada Varela}, {Antoja}, {Baines}, {Baker}, {Balaguer-N{\'u}{\~n}ez}, {Balbinot}, {Balog}, {Barache}, {Barbato}, {Barros}, {Barstow}, {Bartolom{\'e}}, {Bassilana}, {Bauchet}, {Becciani}, {Bellazzini}, {Berihuete}, {Bernet}, {Bertone}, {Bianchi}, {Binnenfeld}, {Blanco-Cuaresma}, {Blazere}, {Boch}, {Bombrun}, {Bossini}, {Bouquillon}, {Bragaglia}, {Bramante}, {Breedt},
  {Bressan}, {Brouillet}, {Brugaletta}, {Bucciarelli}, {Burlacu}, {Butkevich}, {Buzzi}, {Caffau}, {Cancelliere}, {Cantat-Gaudin}, {Carballo}, {Carlucci}, {Carnerero}, {Carrasco}, {Casamiquela}, {Castellani}, {Castro-Ginard}, {Chaoul}, {Charlot}, {Chemin}, {Chiaramida}, {Chiavassa}, {Chornay}, {Comoretto}, {Contursi}, {Cooper}, {Cornez}, {Cowell}, {Crifo}, {Cropper}, {Crosta}, {Crowley}, {Dafonte}, {Dapergolas}, {David}, {David}, {de Laverny}, {De Luise}, {De March}, {De Ridder}, {de Souza}, {de Torres}, {del Peloso}, {del Pozo}, {Delbo}, {Delgado}, {Delisle}, {Demouchy}, {Dharmawardena}, {Di Matteo}, {Diakite}, {Diener}, {Distefano}, {Dolding}, {Edvardsson}, {Enke}, {Fabre}, {Fabrizio}, {Faigler}, {Fedorets}, {Fernique}, {Fienga}, {Figueras}, {Fournier}, {Fouron}, {Fragkoudi}, {Gai}, {Garcia-Gutierrez}, {Garcia-Reinaldos}, {Garc{\'\i}a-Torres}, {Garofalo}, {Gavel}, {Gavras}, {Gerlach}, {Geyer}, {Giacobbe}, {Gilmore}, {Girona}, {Giuffrida}, {Gomel}, {Gomez}, {Gonz{\'a}lez-N{\'u}{\~n}ez},
  {Gonz{\'a}lez-Santamar{\'\i}a}, {Gonz{\'a}lez-Vidal}, {Granvik}, {Guillout}, {Guiraud}, {Guti{\'e}rrez-S{\'a}nchez}, {Guy}, {Hatzidimitriou}, {Hauser}, {Haywood}, {Helmer}, {Helmi}, {Sarmiento}, {Hidalgo}, {Hilger}, {H{\l}adczuk}, {Hobbs}, {Holland}, {Huckle}, {Jardine}, {Jasniewicz}, {Jean-Antoine Piccolo}, {Jim{\'e}nez-Arranz}, {Jorissen}, {Juaristi Campillo}, {Julbe}, {Karbevska}, {Kervella}, {Khanna}, {Kontizas}, {Kordopatis}, {Korn}, {K{\'o}sp{\'a}l}, {Kostrzewa-Rutkowska}, {Kruszy{\'n}ska}, {Kun}, {Laizeau}, {Lambert}, {Lanza}, {Lasne}, {Le Campion}, {Lebreton}, {Lebzelter}, {Leccia}, {Leclerc}, {Lecoeur-Taibi}, {Liao}, {Licata}, {Lindstr{\o}m}, {Lister}, {Livanou}, {Lobel}, {Lorca}, {Loup}, {Madrero Pardo}, {Magdaleno Romeo}, {Managau}, {Mann}, {Manteiga}, {Marchant}, {Marconi}, {Marcos}, {Marcos Santos}, {Mar{\'\i}n Pina}, {Marinoni}, {Marocco}, {Marshall}, {Martin Polo}, {Mart{\'\i}n-Fleitas}, {Marton}, {Mary}, {Masip}, {Massari}, {Mastrobuono-Battisti}, {Mazeh}, {McMillan}, {Messina}, {Michalik},
  {Millar}, {Mints}, {Molina}, {Molinaro}, {Moln{\'a}r}, {Monari}, {Mongui{\'o}}, {Montegriffo}, {Montero}, {Mor}, {Mora}, {Morbidelli}, {Morel}, {Morris}, {Muraveva}, {Murphy}, {Musella}, {Nagy}, {Noval}, {Oca{\~n}a}, {Ogden}, {Ordenovic}, {Osinde}, {Pagani}, {Pagano}, {Palaversa}, {Palicio}, {Pallas-Quintela}, {Panahi}, {Payne-Wardenaar}, {Pe{\~n}alosa Esteller}, {Penttil{\"a}}, {Pichon}, {Piersimoni}, {Pineau}, {Plachy}, {Plum}, {Poggio}, {Pr{\v{s}}a}, {Pulone}, {Racero}, {Ragaini}, {Rainer}, {Raiteri}, {Rambaux}, {Ramos}, {Ramos-Lerate}, {Re Fiorentin}, {Regibo}, {Richards}, {Rios Diaz}, {Ripepi}, {Riva}, {Rix}, {Rixon}, {Robichon}, {Robin}, {Robin}, {Roelens}, {Rogues}, {Rohrbasser}, {Romero-G{\'o}mez}, {Rowell}, {Royer}, {Ruz Mieres}, {Rybicki}, {Sadowski}, {S{\'a}ez N{\'u}{\~n}ez}, {Sagrist{\`a} Sell{\'e}s}, {Sahlmann}, {Salguero}, {Samaras}, {Sanchez Gimenez}, {Sanna}, {Santove{\~n}a}, {Sarasso}, {Schultheis}, {Sciacca}, {Segol}, {Segovia}, {S{\'e}gransan}, {Semeux}, {Shahaf}, {Siddiqui}, {Siebert},
  {Siltala}, {Silvelo}, {Slezak}, {Slezak}, {Smart}, {Snaith}, {Solano}, {Solitro}, {Souami}, {Souchay}, {Spagna}, {Spina}, {Spoto}, {Steele}, {Steidelm{\"u}ller}, {Stephenson}, {S{\"u}veges}, {Surdej}, {Szabados}, {Szegedi-Elek}, {Taris}, {Taylor}, {Teixeira}, {Tolomei}, {Tonello}, {Torra}, {Torra}, {Torralba Elipe}, {Trabucchi}, {Tsounis}, {Turon}, {Ulla}, {Unger}, {Vaillant}, {van Dillen}, {van Reeven}, {Vanel}, {Vecchiato}, {Viala}, {Vicente}, {Voutsinas}, {Weiler}, {Wevers}, {Wyrzykowski}, {Yoldas}, {Yvard}, {Zhao}, {Zorec}, {Zucker}, \& {Zwitter}}]{GAIA2023}
{Gaia Collaboration}, {Vallenari}, A., {Brown}, A.~G.~A., {et~al.} 2023, \aap, 674, A1, \dodoi{10.1051/0004-6361/202243940}

\bibitem[{{Gei{\ss}ler} {et~al.}(2008){Gei{\ss}ler}, {Chauvin}, \& {Sterzik}}]{Geissler2008}
{Gei{\ss}ler}, K., {Chauvin}, G., \& {Sterzik}, M.~F. 2008, \aap, 480, 193, \dodoi{10.1051/0004-6361:20078229}

\bibitem[{{Guenther} {et~al.}(2001){Guenther}, {Neuh{\"a}user}, {Hu{\'e}lamo}, {Brandner}, \& {Alves}}]{Guenther2001}
{Guenther}, E.~W., {Neuh{\"a}user}, R., {Hu{\'e}lamo}, N., {Brandner}, W., \& {Alves}, J. 2001, \aap, 365, 514, \dodoi{10.1051/0004-6361:20000051}

\bibitem[{{Henning}(2010)}]{Henning2010}
{Henning}, T. 2010, \araa, 48, 21, \dodoi{10.1146/annurev-astro-081309-130815}

\bibitem[{{Holman} \& {Wiegert}(1999)}]{Holman1999}
{Holman}, M.~J., \& {Wiegert}, P.~A. 1999, \aj, 117, 621, \dodoi{10.1086/300695}

\bibitem[{{Houk} \& {Cowley}(1975)}]{Houk1975}
{Houk}, N., \& {Cowley}, A.~P. 1975, {University of Michigan Catalogue of two-dimensional spectral types for the HD stars. Volume I. Declinations -90\_ to -53\_{\textflorin}0.}

\bibitem[{{Hughes} {et~al.}(2018){Hughes}, {Duch{\^e}ne}, \& {Matthews}}]{Hughes2018}
{Hughes}, A.~M., {Duch{\^e}ne}, G., \& {Matthews}, B.~C. 2018, \araa, 56, 541, \dodoi{10.1146/annurev-astro-081817-052035}

\bibitem[{Hunter(2007)}]{matplotlib07}
Hunter, J.~D. 2007, Computing In Science \& Engineering, 9, 90, \dodoi{10.1109/MCSE.2007.55}

\bibitem[{{Iglesias} {et~al.}(2023){Iglesias}, {Pani{\'c}}, \& {Rebollido}}]{Iglesias2023}
{Iglesias}, D.~P., {Pani{\'c}}, O., \& {Rebollido}, I. 2023, \mnras, 526, 2500, \dodoi{10.1093/mnras/stad2836}

\bibitem[{{Law} {et~al.}(2023){Law}, {E. Morrison}, {Argyriou}, {Patapis}, {{\'A}lvarez-M{\'a}rquez}, {Labiano}, \& {Vandenbussche}}]{law2023}
{Law}, D.~R., {E. Morrison}, J., {Argyriou}, I., {et~al.} 2023, \aj, 166, 45, \dodoi{10.3847/1538-3881/acdddc}

\bibitem[{{Lazzoni} {et~al.}(2020){Lazzoni}, {Zurlo}, {Desidera}, {Mesa}, {Fontanive}, {Bonavita}, {Ertel}, {Rice}, {Vigan}, {Boccaletti}, {Bonnefoy}, {Chauvin}, {Delorme}, {Gratton}, {Houll{\'e}}, {Maire}, {Meyer}, {Rickman}, {Spalding}, {Asensio-Torres}, {Langlois}, {M{\"u}ller}, {Baudino}, {Beuzit}, {Biller}, {Brandner}, {Buenzli}, {Cantalloube}, {Cheetham}, {Cudel}, {Feldt}, {Galicher}, {Janson}, {Hagelberg}, {Henning}, {Kasper}, {Keppler}, {Lagrange}, {Lannier}, {LeCoroller}, {Mouillet}, {Peretti}, {Perrot}, {Salter}, {Samland}, {Schmidt}, {Sissa}, \& {Wildi}}]{Lazzoni2020}
{Lazzoni}, C., {Zurlo}, A., {Desidera}, S., {et~al.} 2020, \aap, 641, A131, \dodoi{10.1051/0004-6361/201937290}

\bibitem[{{Lowrance} {et~al.}(2000){Lowrance}, {Schneider}, {Kirkpatrick}, {Becklin}, {Weinberger}, {Zuckerman}, {Plait}, {Malmuth}, {Heap}, {Schultz}, {Smith}, {Terrile}, \& {Hines}}]{Lowrance2000}
{Lowrance}, P.~J., {Schneider}, G., {Kirkpatrick}, J.~D., {et~al.} 2000, \apj, 541, 390, \dodoi{10.1086/309437}

\bibitem[{{Mamajek} \& {Bell}(2014)}]{Mamajek2014}
{Mamajek}, E.~E., \& {Bell}, C. P.~M. 2014, \mnras, 445, 2169, \dodoi{10.1093/mnras/stu1894}

\bibitem[{{Marino} {et~al.}(2016){Marino}, {Matr{\`a}}, {Stark}, {Wyatt}, {Casassus}, {Kennedy}, {Rodriguez}, {Zuckerman}, {Perez}, {Dent}, {Kuchner}, {Hughes}, {Schneider}, {Steele}, {Roberge}, {Donaldson}, \& {Nesvold}}]{Marino2016}
{Marino}, S., {Matr{\`a}}, L., {Stark}, C., {et~al.} 2016, \mnras, 460, 2933, \dodoi{10.1093/mnras/stw1216}

\bibitem[{{Milli} {et~al.}(2023){Milli}, {Choquet}, {Tazaki}, {M{\'e}nard}, {Augereau}, {Olofsson}, {Th{\'e}bault}, {Poch}, {Levasseur-Regourd}, {Lasue}, {Renard}, {Hadamcik}, {Baruteau}, {Schmid}, {Engler}, {van Holstein}, {Zubko}, {Lagrange}, {Marino}, {Pinte}, {Dominik}, {Boccaletti}, {Langlois}, {Zurlo}, {Desgrange}, {Gluck}, {Mouillet}, {Costille}, \& {Sauvage}}]{Milli2023}
{Milli}, J., {Choquet}, E., {Tazaki}, R., {et~al.} 2023, arXiv e-prints, arXiv:2312.02000, \dodoi{10.48550/arXiv.2312.02000}

\bibitem[{{Mittal} {et~al.}(2015){Mittal}, {Chen}, {Jang-Condell}, {Manoj}, {Sargent}, {Watson}, \& {Lisse}}]{Mittal2015}
{Mittal}, T., {Chen}, C.~H., {Jang-Condell}, H., {et~al.} 2015, \apj, 798, 87, \dodoi{10.1088/0004-637X/798/2/87}

\bibitem[{{Neuh{\"a}user} {et~al.}(2011){Neuh{\"a}user}, {Ginski}, {Schmidt}, \& {Mugrauer}}]{Neuhauser2011}
{Neuh{\"a}user}, R., {Ginski}, C., {Schmidt}, T.~O.~B., \& {Mugrauer}, M. 2011, \mnras, 416, 1430, \dodoi{10.1111/j.1365-2966.2011.19139.x}

\bibitem[{Nogueira {et~al.}(2024)Nogueira, Lazzoni, Zurlo, Bhowmik, Donoso-Oliva, Desidera, Milli, Pérez, Delorme, Fernadez, Langlois, Petrus, Cabrera-Vives, \& Chauvin}]{nogueira2024astrometric}
Nogueira, P.~H., Lazzoni, C., Zurlo, A., {et~al.} 2024, Astrometric and photometric characterization of $\eta$ Tel B combining two decades of observations.
\newblock \doarXiv{2405.04723}

\bibitem[{Oliphant(2007)}]{numpy07}
Oliphant, T.~E. 2007, Computing in Science \& Engineering, 9

\bibitem[{{Pan} {et~al.}(2016){Pan}, {Nesvold}, \& {Kuchner}}]{Pan2016}
{Pan}, M., {Nesvold}, E.~R., \& {Kuchner}, M.~J. 2016, \apj, 832, 81, \dodoi{10.3847/0004-637X/832/1/81}

\bibitem[{{Patapis} {et~al.}(2024){Patapis}, {Argyriou}, {Law}, {Glauser}, {Glasse}, {Labiano}, {{\'A}lvarez-M{\'a}rquez}, {Kavanagh}, {Gasman}, {Mueller}, {Larson}, {Vandenbussche}, {Lee}, {Klaassen}, {Guillard}, \& {Wright}}]{Patapis2024}
{Patapis}, P., {Argyriou}, I., {Law}, D.~R., {et~al.} 2024, \aap, 682, A53, \dodoi{10.1051/0004-6361/202347339}

\bibitem[{{Pearce} {et~al.}(2024){Pearce}, {Krivov}, {Sefilian}, {Jankovic}, {L{\"o}hne}, {Morgner}, {Wyatt}, {Booth}, \& {Marino}}]{Pearce2024}
{Pearce}, T.~D., {Krivov}, A.~V., {Sefilian}, A.~A., {et~al.} 2024, \mnras, 527, 3876, \dodoi{10.1093/mnras/stad3462}

\bibitem[{{Rameau} {et~al.}(2013){Rameau}, {Chauvin}, {Lagrange}, {Klahr}, {Bonnefoy}, {Mordasini}, {Bonavita}, {Desidera}, {Dumas}, \& {Girard}}]{Rameau2013}
{Rameau}, J., {Chauvin}, G., {Lagrange}, A.~M., {et~al.} 2013, \aap, 553, A60, \dodoi{10.1051/0004-6361/201220984}

\bibitem[{{Rebollido} {et~al.}(2018){Rebollido}, {Eiroa}, {Montesinos}, {Maldonado}, {Villaver}, {Absil}, {Bayo}, {Canovas}, {Carmona}, {Chen}, {Ertel}, {Garufi}, {Henning}, {Iglesias}, {Launhardt}, {Liseau}, {Meeus}, {Mo{\'o}r}, {Mora}, {Olofsson}, {Rauw}, \& {Riviere-Marichalar}}]{Rebollido2018}
{Rebollido}, I., {Eiroa}, C., {Montesinos}, B., {et~al.} 2018, \aap, 614, A3, \dodoi{10.1051/0004-6361/201732329}

\bibitem[{{Rebull} {et~al.}(2008){Rebull}, {Stapelfeldt}, {Werner}, {Mannings}, {Chen}, {Stauffer}, {Smith}, {Song}, {Hines}, \& {Low}}]{Rebull2008}
{Rebull}, L.~M., {Stapelfeldt}, K.~R., {Werner}, M.~W., {et~al.} 2008, \apj, 681, 1484, \dodoi{10.1086/588182}

\bibitem[{{Rigby} {et~al.}(2023){Rigby}, {Lightsey}, {Garc{\'\i}a Mar{\'\i}n}, {Bowers}, {Smith}, {Glasse}, {McElwain}, {Rieke}, {Chary}, {Liu}, {Clampin}, {Kimble}, {Kinzel}, {Laidler}, {Mehalick}, {Noriega-Crespo}, {Shivaei}, {Skelton}, {Stark}, {Temim}, {Wei}, \& {Willott}}]{Rigby2023}
{Rigby}, J.~R., {Lightsey}, P.~A., {Garc{\'\i}a Mar{\'\i}n}, M., {et~al.} 2023, \pasp, 135, 048002, \dodoi{10.1088/1538-3873/acbcf4}

\bibitem[{{Rodet} {et~al.}(2017){Rodet}, {Beust}, {Bonnefoy}, {Lagrange}, {Galli}, {Ducourant}, \& {Teixeira}}]{Rodet2017}
{Rodet}, L., {Beust}, H., {Bonnefoy}, M., {et~al.} 2017, \aap, 602, A12, \dodoi{10.1051/0004-6361/201630269}

\bibitem[{{Sargent} {et~al.}(2009){Sargent}, {Forrest}, {Tayrien}, {McClure}, {Li}, {Basu}, {Manoj}, {Watson}, {Bohac}, {Furlan}, {Kim}, {Green}, \& {Sloan}}]{sargent2009ApJ}
{Sargent}, B.~A., {Forrest}, W.~J., {Tayrien}, C., {et~al.} 2009, \apj, 690, 1193, \dodoi{10.1088/0004-637X/690/2/1193}

\bibitem[{{Schneider} {et~al.}(2006){Schneider}, {Silverstone}, {Hines}, {Augereau}, {Pinte}, {M{\'e}nard}, {Krist}, {Clampin}, {Grady}, {Golimowski}, {Ardila}, {Henning}, {Wolf}, \& {Rodmann}}]{Schneider2006}
{Schneider}, G., {Silverstone}, M.~D., {Hines}, D.~C., {et~al.} 2006, \apj, 650, 414, \dodoi{10.1086/506507}

\bibitem[{{Sefilian}(2024)}]{Sefilian2024}
{Sefilian}, A.~A. 2024, \apj, 966, 140, \dodoi{10.3847/1538-4357/ad32d1}

\bibitem[{{Sefilian} {et~al.}(2021){Sefilian}, {Rafikov}, \& {Wyatt}}]{Sefilian2021}
{Sefilian}, A.~A., {Rafikov}, R.~R., \& {Wyatt}, M.~C. 2021, \apj, 910, 13, \dodoi{10.3847/1538-4357/abda46}

\bibitem[{{Smith} {et~al.}(2009){Smith}, {Churcher}, {Wyatt}, {Moerchen}, \& {Telesco}}]{Smith2009}
{Smith}, R., {Churcher}, L.~J., {Wyatt}, M.~C., {Moerchen}, M.~M., \& {Telesco}, C.~M. 2009, \aap, 493, 299, \dodoi{10.1051/0004-6361:200810706}

\bibitem[{{Stolker} {et~al.}(2020){Stolker}, {Quanz}, {Todorov}, {K{\"u}hn}, {Molli{\`e}re}, {Meyer}, {Currie}, {Daemgen}, \& {Lavie}}]{Stolker2020}
{Stolker}, T., {Quanz}, S.~P., {Todorov}, K.~O., {et~al.} 2020, \aap, 635, A182, \dodoi{10.1051/0004-6361/201937159}

\bibitem[{{Takeuchi} \& {Artymowicz}(2001)}]{Takeuchi2001}
{Takeuchi}, T., \& {Artymowicz}, P. 2001, \apj, 557, 990, \dodoi{10.1086/322252}

\bibitem[{{The Astropy Collaboration} {et~al.}(2018){The Astropy Collaboration}, {Price-Whelan}, {Sip{\H o}cz}, {G{\"u}nther}, {Lim}, {Crawford}, {Conseil}, {Shupe}, {Craig}, {Dencheva}, {Ginsburg}, {VanderPlas}, {Bradley}, {P{\'e}rez-Su{\'a}rez}, {de Val-Borro}, {Paper Contributors}, {Aldcroft}, {Cruz}, {Robitaille}, {Tollerud}, {Coordination Committee}, {Ardelean}, {Babej}, {Bach}, {Bachetti}, {Bakanov}, {Bamford}, {Barentsen}, {Barmby}, {Baumbach}, {Berry}, {Biscani}, {Boquien}, {Bostroem}, {Bouma}, {Brammer}, {Bray}, {Breytenbach}, {Buddelmeijer}, {Burke}, {Calderone}, {Cano Rodr{\'{\i}}guez}, {Cara}, {Cardoso}, {Cheedella}, {Copin}, {Corrales}, {Crichton}, {D'Avella}, {Deil}, {Depagne}, {Dietrich}, {Donath}, {Droettboom}, {Earl}, {Erben}, {Fabbro}, {Ferreira}, {Finethy}, {Fox}, {Garrison}, {Gibbons}, {Goldstein}, {Gommers}, {Greco}, {Greenfield}, {Groener}, {Grollier}, {Hagen}, {Hirst}, {Homeier}, {Horton}, {Hosseinzadeh}, {Hu}, {Hunkeler}, {Ivezi{\'c}}, {Jain}, {Jenness}, {Kanarek}, {Kendrew}, {Kern},
  {Kerzendorf}, {Khvalko}, {King}, {Kirkby}, {Kulkarni}, {Kumar}, {Lee}, {Lenz}, {Littlefair}, {Ma}, {Macleod}, {Mastropietro}, {McCully}, {Montagnac}, {Morris}, {Mueller}, {Mumford}, {Muna}, {Murphy}, {Nelson}, {Nguyen}, {Ninan}, {N{\"o}the}, {Ogaz}, {Oh}, {Parejko}, {Parley}, {Pascual}, {Patil}, {Patil}, {Plunkett}, {Prochaska}, {Rastogi}, {Reddy Janga}, {Sabater}, {Sakurikar}, {Seifert}, {Sherbert}, {Sherwood-Taylor}, {Shih}, {Sick}, {Silbiger}, {Singanamalla}, {Singer}, {Sladen}, {Sooley}, {Sornarajah}, {Streicher}, {Teuben}, {Thomas}, {Tremblay}, {Turner}, {Terr{\'o}n}, {van Kerkwijk}, {de la Vega}, {Watkins}, {Weaver}, {Whitmore}, {Woillez}, {Zabalza}, \& {Contributors}}]{astropy18}
{The Astropy Collaboration}, {Price-Whelan}, A.~M., {Sip{\H o}cz}, B.~M., {et~al.} 2018, \aj, 156, 123, \dodoi{10.3847/1538-3881/aabc4f}

\bibitem[{{Tremaine} {et~al.}(2009){Tremaine}, {Touma}, \& {Namouni}}]{Tremaine2009}
{Tremaine}, S., {Touma}, J., \& {Namouni}, F. 2009, \aj, 137, 3706, \dodoi{10.1088/0004-6256/137/3/3706}

\bibitem[{{Trifonov} {et~al.}(2020){Trifonov}, {Tal-Or}, {Zechmeister}, {Kaminski}, {Zucker}, \& {Mazeh}}]{trifonov2020}
{Trifonov}, T., {Tal-Or}, L., {Zechmeister}, M., {et~al.} 2020, \aap, 636, A74, \dodoi{10.1051/0004-6361/201936686}

\bibitem[{{Vousden} {et~al.}(2016){Vousden}, {Farr}, \& {Mandel}}]{Vousden2016}
{Vousden}, W.~D., {Farr}, W.~M., \& {Mandel}, I. 2016, \mnras, 455, 1919, \dodoi{10.1093/mnras/stv2422}

\bibitem[{{Wells} {et~al.}(2015){Wells}, {Pel}, {Glasse}, {Wright}, {Aitink-Kroes}, {Azzollini}, {Beard}, {Brandl}, {Gallie}, {Geers}, {Glauser}, {Hastings}, {Henning}, {Jager}, {Justtanont}, {Kruizinga}, {Lahuis}, {Lee}, {Martinez-Delgado}, {Mart{\'\i}nez-Galarza}, {Meijers}, {Morrison}, {M{\"u}ller}, {Nakos}, {O'Sullivan}, {Oudenhuysen}, {Parr-Burman}, {Pauwels}, {Rohloff}, {Schmalzl}, {Sykes}, {Thelen}, {van Dishoeck}, {Vandenbussche}, {Venema}, {Visser}, {Waters}, \& {Wright}}]{Wells15}
{Wells}, M., {Pel}, J.~W., {Glasse}, A., {et~al.} 2015, \pasp, 127, 646, \dodoi{10.1086/682281}

\bibitem[{{Wisdom}(1980)}]{Wisdom1980}
{Wisdom}, J. 1980, \aj, 85, 1122, \dodoi{10.1086/112778}

\bibitem[{{Worthen} {et~al.}(2024){Worthen}, {Chen}, {Law}, {Lu}, {Hoch}, {Chai}, {Sloan}, {Sargent}, {Kammerer}, {Hines}, {Rebollido}, {Balmer}, {Perrin}, {Watson}, {Pueyo}, {Girard}, {Lisse}, \& {Stark}}]{Worthen2024}
{Worthen}, K., {Chen}, C.~H., {Law}, D.~R., {et~al.} 2024, \apj, 964, 168, \dodoi{10.3847/1538-4357/ad2354}

\bibitem[{{Wyatt}(2020)}]{Wyatt2020}
{Wyatt}, M. 2020, in The Trans-Neptunian Solar System, ed. D.~{Prialnik}, M.~A. {Barucci}, \& L.~{Young}, 351--376, \dodoi{10.1016/B978-0-12-816490-7.00016-3}

\bibitem[{{Wyatt}(2005)}]{Wyatt2005}
{Wyatt}, M.~C. 2005, \aap, 440, 937, \dodoi{10.1051/0004-6361:20053391}

\bibitem[{{Wyatt} {et~al.}(1999){Wyatt}, {Dermott}, {Telesco}, {Fisher}, {Grogan}, {Holmes}, \& {Pi{\~n}a}}]{Wyatt1999}
{Wyatt}, M.~C., {Dermott}, S.~F., {Telesco}, C.~M., {et~al.} 1999, \apj, 527, 918, \dodoi{10.1086/308093}

\bibitem[{{Youngblood} {et~al.}(2021){Youngblood}, {Roberge}, {MacGregor}, {Brandeker}, {Weinberger}, {P{\'e}rez}, {Grady}, \& {Welsh}}]{Youngblood2021}
{Youngblood}, A., {Roberge}, A., {MacGregor}, M.~A., {et~al.} 2021, \aj, 162, 235, \dodoi{10.3847/1538-3881/ac21d1}

\end{thebibliography}


\newpage
\appendix
\section{Corner plots}
\label{app:1} 

\begin{figure*}[hbp]
    \centering
    \includegraphics[width=\textwidth]{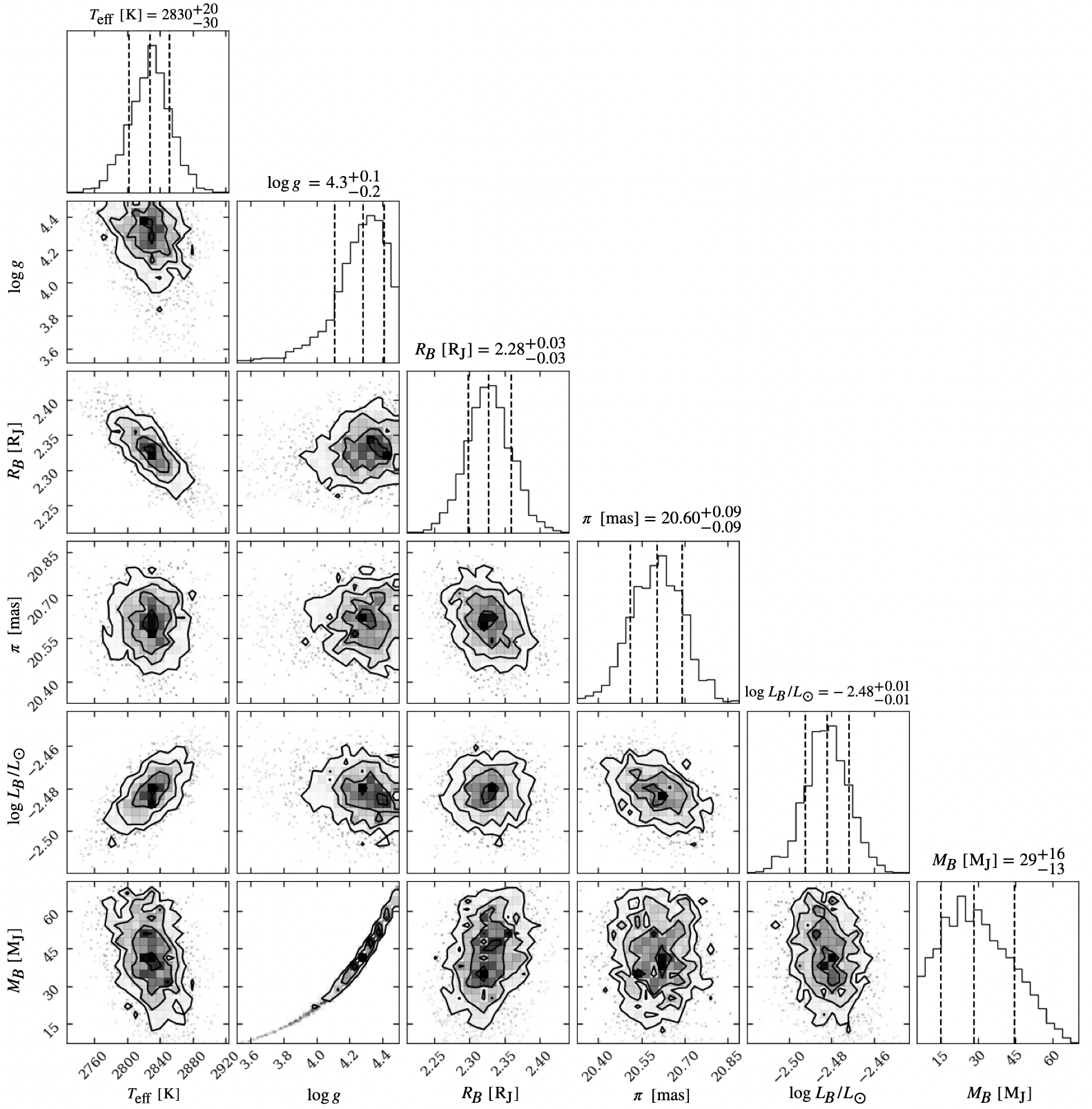}
    \caption{Corner plot for best-fit properties of $\eta$ Tel B derived from a BT-SETTL (CIFIST) model grid fit using \texttt{species} (Stolker et al. 2020)}
    \label{fig:14}
\end{figure*}

\begin{figure}[h!]
    \centering
    \includegraphics[width=\textwidth]{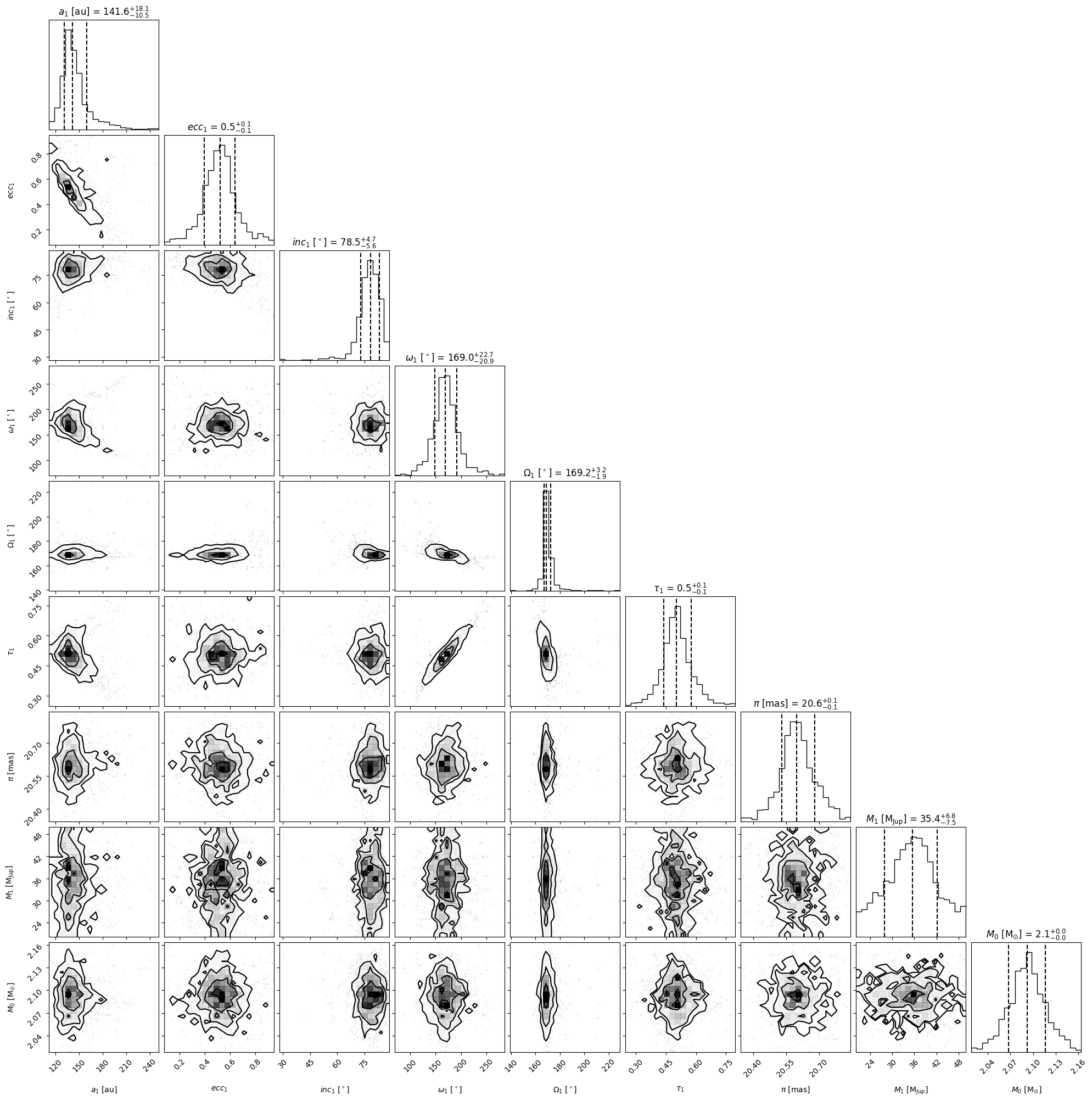}
    \caption{Corner plot for derivation of $\eta$ Tel B orbital parameters using \texttt{orbitize!} \citep{Blunt2020}, as discussed in \hyperref[ssec:Borbfit]{Section 5.2}.}
    \label{fig:15}
\end{figure}

\section{Additional \texttt{orbitize!} fits for $\eta$ Tel B}
\begin{table}[hbp]
    \begin{center}
        \caption{Summary of Median \texttt{orbitize!} Posteriors for Different Diagnostic Fits}
        \begin{tabular}{cccccc}
            \hline
            Model Parameter&  \hyperref[ssec:Borbfit]{Section 5.2} Fit&  N24 Companion Mass Prior& N24 Relative Astrometry Only& N24 MCMC Initial Positions& Relaxed Primary Mass Prior\\
            \hline
            $a_B$ [au]&  142$^{+18}_{-11}$&  141$^{+15}_{-9}$& 140$^{+13}_{-8}$& 149$^{+79}_{-13}$& 142$^{+14}_{-9}$\\
            $e_B$&  0.5$\pm$0.1& 0.5$\pm$0.1& 0.5$\pm$0.1& 0.5$^{+0.1}_{-0.2}$&  0.5$\pm$0.1\\
            $i_B$ [$^\circ$]&  79$^{+5}_{-6}$& 80$^{+5}_{-7}$& 80$^{+5}_{-6}$& 79$^{+6}_{-7}$& 80$\pm$5\\
            $\omega_B$ [$^\circ$]&  169$^{+23}_{-21}$& 170$^{+19}_{-20}$& 171$\pm$19& 34$^{+34}_{-11}$& 170$^{+16}_{-21}$\\
            $\Omega_B$ [$^\circ$]&  169$^{+3}_{-2}$& 169$^{+3}_{-2}$& 169$^{+3}_{-2}$& 344$^{+2}_{-172}$& 169$^{+3}_{-2}$\\
            $\tau_B$&  0.5$\pm$0.1& 0.5$\pm$0.1& 0.5$\pm$0.1& 0.6$^{+0.01}_{-0.4}$& 0.5$\pm$0.1\\
            $M_B$ [$M_J$]&  35$^{+7}_{-8}$& 48$^{+6}_{-7}$& 35$\pm$0.4& 46$^{+12}_{-21}$& 35$\pm$0.6\\
            \hline
        \end{tabular}
    \end{center}
    \vspace{-2mm}
    \footnotesize{Note: uncertainties indicate 68\% ranges. Priors for all parameters are the same as described in \autoref{tab:2}, except for the use of a Gaussian companion mass prior in the N24 Companion Mass Prior and N24 MCMC Initial Positions fits. Initial walker position distributions are all uniform (as described in \hyperref[ssec:Batmomod]{Section 4.1}), except for the N24 MCMC Initial Positions fit, which uses a lognormal distribution for $a$ and a normal distribution for all other parameters.}
    \label{tab:3}
\end{table}

\begin{figure}[hbp]
    \centering
    \includegraphics[width=\textwidth]{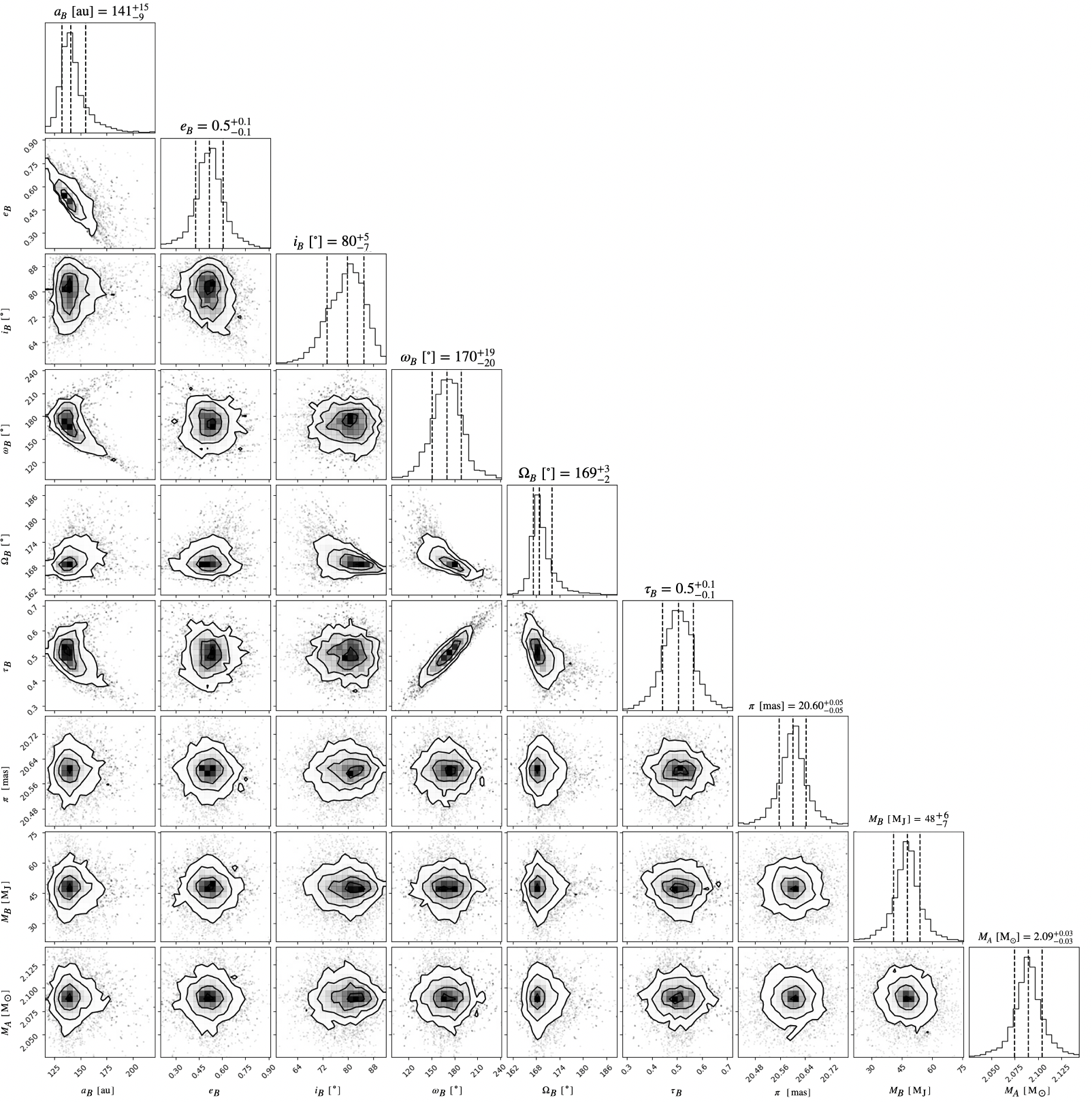}
    \caption{Corner plot for \texttt{orbitize!} \citep{Blunt2020} derivation of $\eta$ Tel B orbital parameters using \cite{nogueira2024astrometric} companion mass Gaussian prior of $M=47\pm0.15 M_J$.}
    \label{fig:16}
\end{figure}

\begin{figure}[hbp]
    \includegraphics[width=\textwidth]{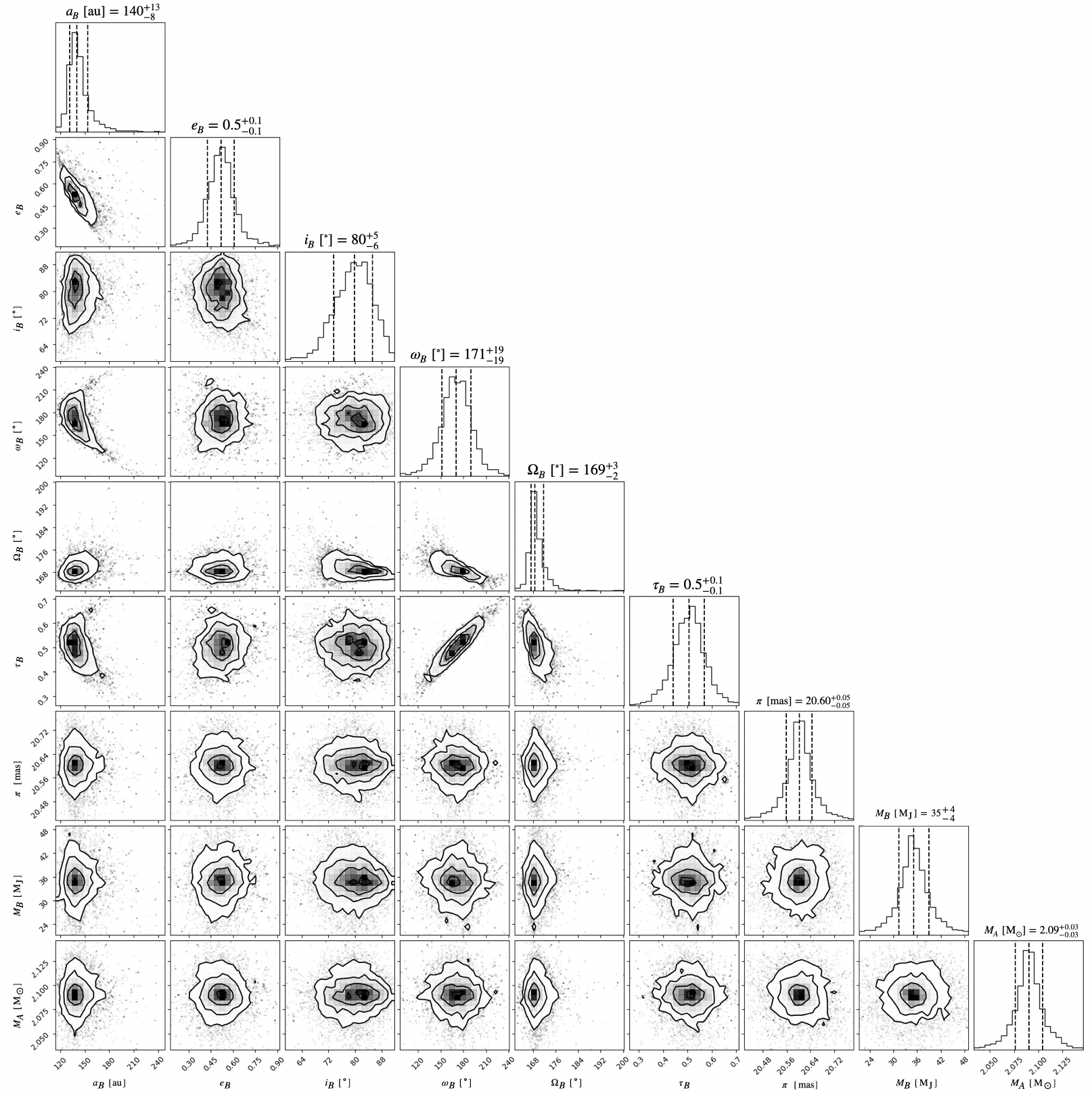}
    \caption{Corner plot for \texttt{orbitize!} \citep{Blunt2020} derivation of $\eta$ Tel B orbital parameters using only relative astormetry included in \cite{nogueira2024astrometric} \texttt{orvara} fit.}
    \label{fig:17}
\end{figure}

\begin{figure}[hbp]
    \includegraphics[width=\textwidth]{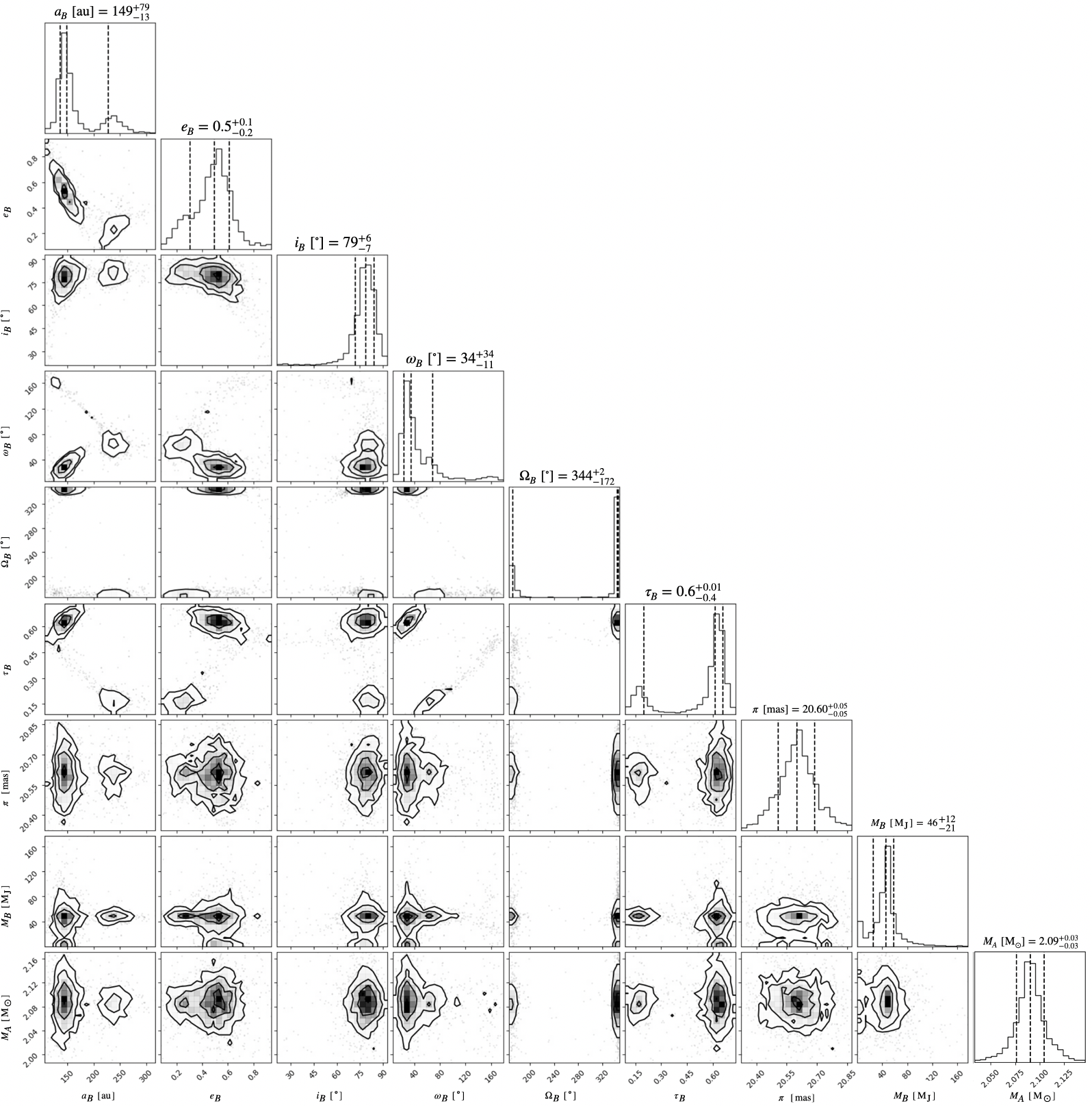}
    \caption{Corner plot for \texttt{orbitize!} \citep{Blunt2020} derivation of $\eta$ Tel B orbital parameters using \cite{nogueira2024astrometric} initial distribution positions and priors.}
    \label{fig:18}
\end{figure}

\begin{figure}[hbp]
    \includegraphics[width=\textwidth]{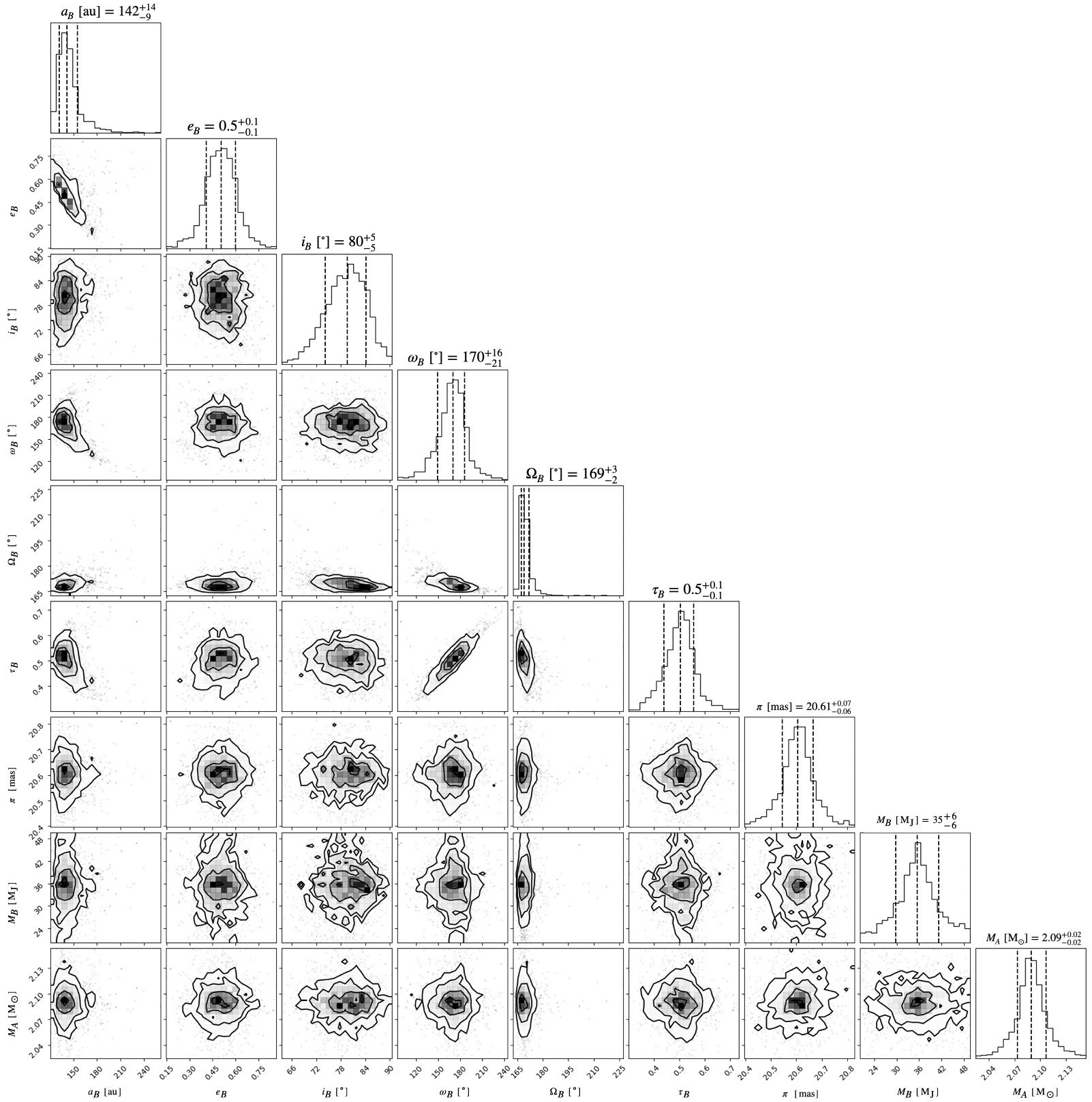}
    \caption{Corner plot for \texttt{orbitize!} \citep{Blunt2020} derivation of $\eta$ Tel B orbital parameters a relaxed primary mass prior of $M=2.2\pm0.1$ \citep{Chen14}.}
    \label{fig:19}
\end{figure}

\end{document}